\documentclass[prc,aps,preprint,showpacs,nofootinbib,showkeys,eqsecnum,tightenlines]{revtex4}    % for hep-ph
\usepackage{epsfig}
\def\be{\begin{equation}}
\def\ee{\end{equation}}
\def\bea{\begin{eqnarray}}
\def\eea{\end{eqnarray}}
\def\gn{\gamma_\nu}
\def\g5{\gamma_5}

\def\vep{\varepsilon}
\def\ot{(1\,\leftrightarrow\,2)}
\def\sa{\vec \sigma_1}
\def\sb{\vec \sigma_2}

\def\qa{\vec q_1}
\def\qb{\vec q_2}
\def\q2{\vec q_2}
\def\vq{\vec q}
\def\Pa{\vec P_1}
\def\Pb{\vec P_2}

\def\s2q2{(\vec \sigma_2 \times \vec q_2)}
\def\ta{\,\tau^{\,a}_1}
\def\tb{\,\tau^{\,a}_2}
\def\t1t2a{\,i\,(\vec \tau_1 \times \vec \tau_2)^a}
\def\tb3{\,\tau^{\,3}_2}
\def\t1t23{\,i\,(\vec \tau_1 \times \vec \tau_2)^3}
\def\tba{\,\tau^{\,a}_2}
\def\t1t2a{\,i\,(\vec \tau_1 \times \vec \tau_2)^a}
\def\fot{\frac{1}{2}}
\newcommand{\noi}{\noindent}

\def\Journal#1#2#3#4{{#1}{\bf #2} (#4) #3 }
\def\NPA{{ Nucl. Phys.} \bf A}
\def\NPB{{ Nucl. Phys.} \bf B}
\def\PRL{ Phys. Rev. Lett.\,\,}
\def\PRC{{ Phys. Rev.} \bf C}
\def\PRD{{ Phys. Rev.} \bf D}
\def\PRW{ Phys. Rev.\,\,}
\def\FBS{ Few--Body Systems\,\,}
\def\PLB{{ Phys. Lett.} \bf B}

\def\EPJA{{ Eur. Phys. J.}\bf A}

\def\PR{ Phys. Rep.\,\,}
\def\IJMPA{{ Int. J. Mod. Phys.} \bf A}
\def\IJMPE{{ Int. J. Mod. Phys.} \bf E}

\def\SJPN{ Sov. J. Part. Nucl.\,\,}
\def\APNY{ Ann. Phys. (N.Y.)\,\,}
\def\CPNP{ Comments Part. Nucl. Phys.\,\,}
\def\ARNPS{ Ann. Rev. Nucl. Part. Sci.\,\,}
\def\NC{ Nuovo Cim.\,\,}
\def\RMP{ Rev. Mod. Phys.\,\,}
\def\ANP{ Adv. Nucl. Phys.\,\,}
\def\JPG{ J. Phys. \bf G}
\def\AJ{ Astrophys. J.\,\,}
\def\CJPB{{ Czech. J. Phys.} \bf B}
\def\PPNP{Prog. Part. Nucl. Phys.\,\,}
\def\HPHA{Helv. Phys. Acta\,\,}
\def\CPC{Comp. Phys. Comm.\,\,}

\begin{document}

\title{
{\large{\bf Weak axial nuclear heavy meson exchange currents and
interactions of solar neutrinos with deuterons}}}
\author{B.~Mosconi}
\affiliation{Universit$\grave{a}$ di Firenze, Department of
Physics, and Istituto Nazionale di Fisica Nucleare, Sezione di
Firenze, I-50019, Sesto Fiorentino (Firenze), Italy }
\author{P. Ricci}
\affiliation{Istituto Nazionale di Fisica Nucleare, Sezione di
Firenze, I-50019, Sesto Fiorentino (Firenze), Italy }
\author{E.~Truhl\'{\i}k}
\affiliation{Institute of Nuclear Physics ASCR, CZ--250 68
\v{R}e\v{z}, Czech Republic}

\begin{abstract}
\noi Starting from the axial heavy meson exchange currents,
constructed earlier in conjunction with the Bethe--Salpeter
equation, we first present the axial $\rho$--, $\omega$-- and
$a_1$ meson exchange Feynman amplitudes that satisfy the partial
conservation of the axial current. Employing these amplitudes, we
derive the corresponding weak axial heavy meson exchange currents
in the leading order in the $1/M$ expansion ($M$ is the nucleon
mass), suitable for the nuclear physics calculations beyond the
threshold energies and with  wave functions obtained by solving
the Schr\"odinger equation with one--boson exchange potentials.
The constructed currents obey the nuclear form of the partial
conservation of the axial current. We apply the space component of
these currents in calculations of the cross sections for the
disintegration of deuterons by low energy (anti)neutrinos. The
deuteron and the final state nucleon--nucleon wave functions are
derived (i) from a variant of the OBEPQB potential, and (ii) from
the Nijmegen 93 and Nijmegen I nucleon-nucleon interaction. The
extracted values of the constant $L_{1,\,A}$, entering the axial
exchange currents of the pionless effective field theory, are in a
reasonable agreement with its value predicted by the dimensional
analysis.
\end{abstract}

\noi


\noi

\hskip 1.9cm 

\maketitle

%%%%%%%%%%%%%%%%%%%%%%%%%%%%%%%%%%%%%%%%%%%%%%%%%%%%%%%%%%%%%%%%%%%%%%%

\section{Introduction}
\label{rintro}

Since quarks are confined, the quantum chromodynamics
\cite{DGH,W2} is not directly suitable for the investigation of
nuclear physics phenomena at low and intermediate energies.
Instead of quarks and gluons one employs effective degrees of
freedom (hadrons), and ideas and methods based on the concept of
the spontaneously broken chiral symmetry \cite{AD,AFFR}. The
hadronic degrees of freedom, relevant for describing a nucleus and
its response to the external electroweak interactions, are
nucleon, $\Delta(1236)$ isobar and low lying mesons at the hadron
mass scale, like $\pi$-, $\rho$-, $\omega$-,... mesons. The
interaction of the vector mesons with the baryons and pions is
fixed by the vector dominance model (VDM) \cite{JJS}. This concept
has been developed successfully during the last 4 decades
\cite{BS,RW,RB,EW,JDW0,CSc}, and is called by some authors as the
Standard Nuclear Physics Approach (SNPA) \cite{NSAPMGK}.

Starting from the early 1970s, a particular effort was devoted to
the study of mesonic degrees of freedom in nuclei by investigating
meson exchange currents (MECs) effects
\cite{CR,KDR,ITEC,ISTAR,KTEC,ISTPR,JFM,DOR,FM}. One of the best
proofs of  presence of the pionic degrees of freedom in nuclei
follows from the study of the transition $0^+\leftrightarrow 0^-$
in the A=16 nuclei \cite{ISTAR,FM}, induced by the time component
of the axial current. Detailed studies have shown that the
experimental data and the calculations can be reconciled only if
the time component of the weak axial soft pion exchange current
\cite{KDR} is taken into account. The hard pion corrections
\cite{JKT,NKO} change the result by 10-15 \%.

The time component of the weak axial MECs plays an important role
also in interpreting the data on the isovector $0^+\leftrightarrow
1^+$ transition in the A=12 nuclei \cite{GS} and on the
$0^+\leftrightarrow 0^-$ transitions in medium--heavy and heavy
nuclei. In these last transitions, the part of this component
arising from the heavy meson exchanges, contributes sizeably
\cite{KRT,To1,EX}. The weak axial exchange charge densities,
derived for the phenomena at the threshold, are given by pair
terms that are related to nucleon--nucleon potentials \cite{KRT},
or are obtained from the chiral Lagrangian \cite{To1}.

In the classification of Ref.\,\cite{KDR}, the leading term of the
space component of the axial MECs of the pion range is of the
order $\sim {\cal O}(1/M^2)$, where $M$ is the nucleon mass. Being
of relativistic origin, it is model dependent. This component of
the weak axial MECs plays an important role in
such fundamental reactions as \\
(i) the neutrino reactions in nuclei at low and intermediate energies, \\
(ii) the weak transitions in light nuclei, in particular the
tritium beta decay, ordinary muon capture in $^{2}$H and $^{3}$He
\footnote{For the recent review on these two reactions see
Refs.\,\cite{My}--\cite{TCJPB}.}, and the solar fusion pp and
p$^{3}$He processes, so important for the determination of the
flux of the solar neutrinos,  \\
(iii) and the parity violating (PV) electron scattering that aims
at elucidating the strange quark contribution to the
electromagnetic structure of the nucleon \footnote{Several
calculations of the PV inclusive \cite{HH,HHM,HPD,GIP,DSK} and
exclusive \cite{MRI,KA} electron scattering off deuterons have
already been done considering different theoretical issues,
however, the axial exchange currents have not yet been included in
the calculations.}.

With the use of the chiral Lagrangians, the structure of the space
component of the weak axial MECs of the pion range was studied in
detail in the hard pion model \cite{IT1} in
Refs.\,\cite{ISTPR,AHHST,TK1}, and applied to various reactions in
the lightest nuclei \cite{AHHST,HPA,CT,TKK}. The largest effect
arises from the $\Delta$ excitation current of the pion range,
that is partially compensated by the analogous current of the
$\rho$ meson range. Besides, the pair current of the $\rho$ meson
range was used during the last decade in
Refs.\,\cite{NSAPMGK,Sch14}. However, the origin of its derivation
remains unclear.

The method used in Ref.\,\cite{KRT} for the construction of the
axial charge at the threshold was applied in Ref.\, \cite{TR} to
the derivation of the space component of the weak axial potential
MECs.

The weak axial MECs of the $N\pi\sigma\omega$ system have recently
been studied in Ref.\,\cite{SMA}, employing the Lagrangian based
on the linear $\sigma$ model \cite{SIM}. The model suffers from
some problems discussed in detail in Ref.\,\cite{ASW}.

The weak axial MECs of the pion range in the Bethe--Salpeter
approach to the nuclear two--body problem has been studied in
various chiral models in Ref.\,\cite{VD}. Another recent
construction of the weak axial  one--boson exchange currents for
the Bethe--Salpeter equation has been done in Ref.\,\cite{KT1}
making use of the chiral Lagrangians of the $N\Delta \pi \rho
\omega a_1$ system \cite{IT1,STG}. The obtained current operators
fulfill the Ward--Takahashi identities and the matrix element of
the full current, sandwiched between the two--body solutions of
the Bethe--Salpeter equation, satisfies the PCAC constraint.

The chiral Lagrangians \cite{IT1,STG} are constructed in such a
way \cite{AFFR,IT1,SIM,STG,NLL,YML,OZ,HLSLM,BKY} that they
reproduce results obtained from the current algebra and PCAC in
the tree approximation. Besides possessing the chiral symmetry,
our Lagrangians are characterized by the following properties: (i)
They respect VDM, reproduce universality, KSFRI, KSFRII. (ii) They
provide the correct anomalous magnetic moment of the $a_1$ meson.
(iii) They reproduce the current algebra prediction for the weak
pion production amplitude. It was explicitly shown \cite{STG} for
the Lagrangian, based on the hidden local  symmetry approach that
if the $a_1$ meson field is eliminated, the resulting Lagrangian
preserves the properties (i) and (iii). Moreover, if the $\rho$
meson field is eliminated from this Lagrangian, the chiral
Lagrangian of the nucleons and pions is recovered \cite{ST}. Hence
our Lagrangians consistently combine the chiral approach with the
VDM concept, and provide a reasonable approximation at the tree
level to the hadron amplitudes up to the energy scale
$\approx\,1\,GeV$ ($\approx\,m_\rho,\,m_{a_1},\,M$). Subsequently,
these Lagrangians were applied to the construction of the weak
axial MECs in the tree approximation: the generic relativistic
Feynman tree-level amplitudes satisfy the PCAC constraint and the
nuclear MECs derived from them are required to satisfy the nuclear
PCAC constraint (see below). In this approach, the weak hadron
form factors are naturally of the VDM form, but the strong nucleon
form factors should be introduced by hand.

In practical calculations, one makes the non--relativistic
reduction of the currents by expanding in $Q/M_h$, where $Q$ is
the momentum of the external particles or the momentum transfer,
and $M_h$ is the heavy meson or nucleon mass. Besides the leading
order terms, the leading relativistic corrections have been
calculated in the electromagnetic sector for the MECs of the pion
range \cite{ATA,GOA}. Due to the presence of the small expansion
parameter, the model currents are expected to be valid in the
energy/momentum region up to $\approx\,0.4$ $GeV$. Let us note
that the cross sections for the backward deuteron
electrodisintegration calculated in \cite{STG} describe well the
data up to $Q^2\,\approx\,1.2$ $(GeV/c)^2$ \cite{SCK,GIG}.

We shall call the above described approach, where the currents and
potentials are constructed in the tree approximation, as Tree
Approximation Approach (TAA). The advantage of this approach is a
relative simplicity and transparency. On the other hand, the
obtained results can be considered  fully realistic. Moreover, the
nuclear PCAC constraint connects a part of the axial nuclear MECs
(potential MECs) with the nuclear potential derived within the
same approach. Making use of this potential in the production of
the nuclear wave functions, one can do consistent calculations of
the potential MECs effect. Let us note that the problem with the
consistency of the calculations is by most authors overlooked
\cite{MRT1}.

In principle, the TAA is improved by effective field theories
(EFTs). An EFT is based on the most general Lagrangian involving
the relevant degrees of freedom, and respecting chiral symmetry
\cite{DGH,W2,W3}. For the pion--nucleon system, this approach,
that was developed intensively in the 1990s \cite{BKM,HHK,KSW},
inspired a burst of applications in the region of low energies
\cite{AHBCHPT}. In particular, the time component of the weak
axial MECs of the pion--nucleon system was constructed \cite{PMR}
within the framework of the heavy baryon chiral perturbation
theory. Besides the tree approximation, it contains the
contribution from the one--loop graphs. It was subsequently
applied \cite{PTK} to the calculation of the MECs effect in nuclei
that has already been discussed above.

The space component of the weak axial MECs of the pion--nucleon
system was constructed within the same scheme in Ref.\,\cite{PKMR}
and applied to the weak interaction processes in the lightest
nuclei in Refs.\,\cite{PKMR,ASP,APKM,ASPFK}. These calculations
are hybrid, since the MECs are taken from the EFT, whereas the
nuclear wave functions are derived from the potential models of
the SNPA. Moreover, as it has recently been discussed in
\cite{MRT1}, the long--range part of these MECs does not satisfy
the nuclear PCAC constraint.

Another EFT \cite{KSW} was applied to the construction of the
space component of the weak axial MECs in Refs.\,\cite{MB1,MB2}
where also its influence on the reactions of the low energy
(anti)neutrinos in deuterons was investigated.

The validity of the EFTs, based on the nucleonic and pionic
degrees of freedom and with the heavy meson and $\Delta$ isobar
degrees of freedom integrated out, is necessarily restricted to
the long wave--length limit  and internuclear distances $r\ge 0.6$
fm \cite{PMR}. It is natural that in parallel with the development
and applications of these EFTs, attempts appeared \cite{MeS} to
construct a class of EFTs without this restriction. The $\Delta$
isobar has already been included explicitly within the small scale
expansion scheme \cite{HHK} and it has recently been demonstrated
in Ref\,.\cite{FSGS} that under certain conditions one can obtain
a consistent power counting for model Lagrangians including vector
mesons.

As it is seen from the discussion, a systematic derivation of the
weak axial heavy MECs operator that would respect chiral symmetry
and VDM is lacking. In the absence of an EFT including explicitly
heavy mesons, we construct here such an operator from our chiral
Lagrangians  in the TAA. Using this operator in calculations of
observables at low energies and comparing the results with the
calculations based on existing EFTs and with the data can provide
a test of soundness of our approach\footnote{A test of this kind
has recently been done in Ref.\,\cite{NSAPMGK}.}.

The main goal of this study should be seen in the construction of
the weak axial nuclear exchange currents (WANECs) of the heavy
meson range, suitable in the SNPA calculations beyond the long
wave--length limit, with the nuclear wave functions generated from
the Schr\"odinger equation and the related one--boson exchange
potentials (OBEPs). For the construction of the WANECs we make use
of the weak axial two--nucleon relativistic amplitudes derived in
\cite{KT1} to which we add the nucleon Born terms. The WANECs are
then defined by analogy with the electromagnetic MECs
\cite{ATA,RI}, as the difference between these relativistic
amplitudes and the first Born iteration of the weak axial
one--nucleon current contribution to the two--nucleon scattering
amplitude satisfying the Lippmann--Schwinger equation. This method
has already been applied in \cite{AHHST,TK1,MRT1} to the
construction of the weak axial MECs of the pion range. It  can be
shown in the same manner that the WANECs, defined in this way,
satisfy the nuclear PCAC equation of the type \cite{ATS} \be q_\mu
A^a_\mu(2)\,=\,[V,\,A^a_0(1)]\,+\,if_\pi
m^2_\pi\Delta^\pi_F(q^2){\cal M}^a(2)\,,  \label{WPCAC} \ee which
follows, as is shortly discussed in appendix \ref{happA}, from the
assumption that the axial current consists of one- and two--body
terms and it satisfies the PCAC hypothesis for the total axial
current. Therefore, these currents are suitable for calculations
of observables for weak processes in the intermediate energy
region exactly as the vector exchange currents satisfying the
nuclear conserved vector current (CVC) equation \be q_\mu
V^a_\mu(2)\,=\,[V,\,V^a_0(1)]\,, \label{NCVC} \ee are applied in
the analogous electromagnetic transitions. The most favorable
situations appear in light nuclei, where the approximation of free
nucleons used for construction of the transition operators is
commonly accepted. In other words, we treat the vector and axial
currents on equal footing and we consider Eq.\,(\ref{WPCAC}) to be
as important for the axial current as Eq.\,(\ref{NCVC}) is
important for the vector current.

The structure of the paper is as follows. In Sect.\,\ref{CH1}, we
consider the two--nucleon weak axial relativistic amplitudes of
the $\rho$-, $\omega$-, and $a_1$ ranges, derived from the
Lagrangian \cite{IT1} and we list the PCAC equations which the
amplitudes satisfy. In Sect.\,\ref{CH2}, we first define the
WANECs as the difference of the relativistic exchange amplitude
and of the first Born iteration of the nuclear equation of motion.
Next we derive the nuclear PCAC that the WANECs should satisfy.
Then we proceed to investigate the structure of the WANECs and
present the resulting currents in the leading order in $1/M$.

In Sect.\,\ref{CH3}, we provide numerical estimates of the cross
sections and of the strength of various parts of the space
component of our WANECs for the weak deuteron disintegration by
the low energy (anti)neutrinos and compare them with the
calculations of Refs.\,\cite{NSAPMGK,MB2,NSGK,YHH}. We discuss our
results in Sect.\,\ref{CH4}. Our notations and basic definitions
are shortly presented in appendix \ref{happA}, and the structure
of the weak axial $\sigma$ meson exchange current is shortly
discussed in appendix \ref{happB}.

\section{Two--nucleon weak axial meson exchange amplitudes of the $\rho$, $a_1$ and $\omega$
 ranges  \label{CH1}}

We first write the weak axial amplitude for the i{\it th} nucleon
(i$\,=\,1,\,2$) \be
J^a_{\,5\mu}(1,i)\,=\,-i\frac{g_A}{2}\,m^2_{a_1}\,\Delta_{\mu\,\nu}^{a_1}(q_i)\,
 {\tilde \Gamma}^{a}_{5\,\nu}(i)\,
-i\,f_\pi\,q_{i\,\mu}\,\Delta^\pi_F(q^2_i)\,\Gamma^{\pi
a}_i\,\equiv\, \bar{u}(p'_i)\,{\hat
J}_{\,5\mu}(1,i)\frac{\tau^a_i}{2}\,u(p_i). \label{ONAC} \ee Here
q$_{\,i}\,=\,$p$'{\,_i}\,-\,$p$_{\,i}$, the vector-meson
propagator is generally designed as \be
\Delta_{\mu\,\nu}^B(q)\,=\,\left(\delta_{\mu\,\nu}\,+\,\frac{q_\mu\,q_\nu}{m^2_B}
\right)\,\Delta^B_F(q^2),
\quad\Delta^B_F(q^2)\,=\,\frac{1}{m^2_B\,+\,q^2}\,,  \label{prop}
\ee and the pseudovector and pseudoscalar vertices are defined as
\be {\tilde \Gamma}^{a}_{5\,\nu}(i)\,=\,\bar{u}(p'_i)\, \gn \g5
\tau^a_i\, u(p_i)\,,\quad \Gamma^{\pi
a}_i\,=\,ig\,\bar{u}(p'_i)\,\g5 \tau^a_i\,u(p_i)\,\equiv\,
\bar{u}(p'_i)\,\hat{\cal O}^\pi_i\tau^a_i\,u(p_i)\,.
\label{vpvps} \ee The divergence of the amplitude (\ref{ONAC}) is
\be q_\mu J^a_{\,5\mu}(1,i)\,=\,i\,f_\pi
m^2_\pi\,\Delta^\pi_F(q^2_i)\,
                  M^a(1,i)\,. \label{DOBCi}
\ee Here  the one--body pion absorption amplitude $M^a(1,i)$ is
defined as \be M^a(1,i)\,=\,\Gamma^{\pi a}_i\,. \label{def1} \ee

\begin{figure}[h!]
\centerline{ \epsfig{file=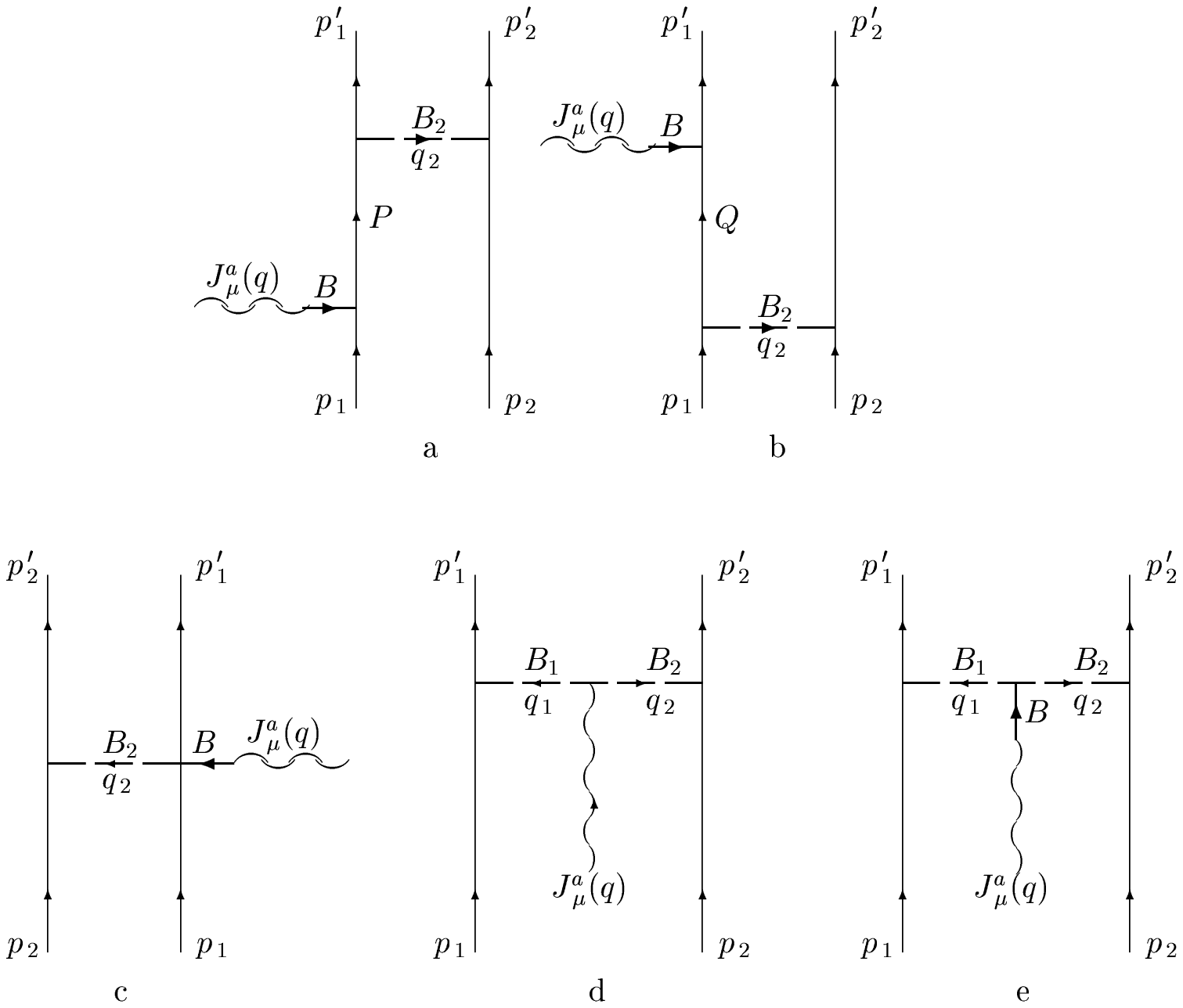} } \vskip 0.4cm \caption{ The
general structure of the two--nucleon weak axial amplitudes
considered in this paper. The weak axial interaction is mediated
by the meson B that is either $\pi$ or $a_1$ meson. The range of
the amplitude is given by the meson $B_2$ that is here $\rho$,
$a_1$ or $\omega$ meson; (a),(b) -- the nucleon Born amplitude
${J}^a_{5\mu,\,B_2}(B)$; (c) -- a contact amplitude ${
J}^a_{5\mu,\,c\,B_2}(B)$; it is connected with the weak production
amplitude of the $B_2$ meson on the nucleon. Another type of the
contact terms, ${ J}^a_{5\mu,\,B_1\,B_2}$, is given in (d), where
the weak axial amplitude interacts directly with the mesons $B_1$
and $B_2$. (e) -- a mesonic amplitude ${
J}^a_{5\mu,\,B_1\,B_2}(B)$. The associated pion absorption
amplitudes correspond to the graphs where the weak axial
interaction is mediated by the pion, but with the weak interaction
wavy line removed. There are three types of these amplitudes in
our models: $ { M}^a_{B_2}$, $ { M}^a_{c\,B_2}$ and ${
M}^a_{B_1\,B_2}$. } \label{figg1}
\end{figure}

The general structure of the two--nucleon weak axial amplitudes of
our model is given in Fig.\,\ref{figg1}.  We shall next consider
$\rho$, $a_1$ and $\omega$ exchanges. As in the pion exchange case
\cite{AHHST,TK1}, we study first the relativistic exchange
amplitudes.

\subsection{Two--nucleon weak axial exchange amplitudes of the $\rho$,
$a_1$ and $\omega$ meson range \label{CH12}}

Let us first write down the general form of the nucleon Born
amplitude \bea J^a_{\,5\mu,\,B}&\equiv&
J^a_{\,5\mu,\,B}(a_1)+J^a_{\,5\mu,\,B}(\pi)=-\bar{u}(p'_1)\left[\,
\hat {\cal O}^B_{1(\nu)}(-q_2)S_F(P)\,{\hat
J}_{\,5\mu}(1,q)\frac{1}{2}(a^+ -a^-)+
{\hat J}_{\,5\mu}(1,q)\right. \nonumber \\
&& \left. \times\,S_F(Q)\hat {\cal
O}^B_{1(\nu)}(-q_2)\frac{1}{2}(a^+ +a^-) \right]u(p_1)
\Delta^B_{(\nu\eta)}(q_2)\bar{u}(p'_2)\hat {\cal
O}^B_{2(\eta)}(q_2) u(p_2)+ \ot\,,  \label{JbBf} \eea where it
holds for various meson exchanges \bea B&=&\rho\,,\quad\hat {\cal
O}^\rho_{i\eta}(q_2)=-i\frac{g_\rho}{2}(\gamma_\eta-
\frac{\kappa^V_\rho}{2M}\sigma_{\eta\delta} q_{2\delta})_i \equiv
-ig_{\rho NN}
\tilde {\cal O}^\rho_{i\eta}(q_2)\,, \label{Orho} \\
B&=&\omega\,,\quad\hat {\cal
O}^\omega_{i\eta}(q_2)=-i\frac{g_\omega}{2}(\gamma_\eta-
\frac{\kappa^S}{2M}\sigma_{\eta\delta} q_{2\delta})_i \equiv
-ig_{\omega NN}\tilde {\cal O}^\omega_{i\eta}(q_2)\,,
\quad \tau^n\rightarrow 1\,, \label{Oomega} \\
B&=&a_1\,,\quad\hat {\cal O}^{a_1}_{i\eta}=-ig_\rho
g_A(\gamma_\eta\gamma_5)_i
\equiv -ig_{a_1 NN}\tilde {\cal O}^{a_1}_{i\eta}\,, \label{Oa1} \\
\eea and \be
a^\pm\,=\,\frac{1}{2}\,[\tau^a_1,\,\tau^n_1]_\pm\,\tau^n_2\,.
\label{apm} \ee For the isovector meson exchange \be a^+\,=\,
\tau^a_2\,,\quad a^-\,=\,-i(\vec \tau_1 \times \vec \tau_2)^a\,,
\label{ivb} \ee whereas for the isoscalar meson exchange \be
a^+\,=\, \tau^a_1\,,\quad a^-\,=\,0\,. \label{isb} \ee As it will
become clear soon, the $\rho$ and $a_1$ exchanges should be
considered in the  chiral model  \cite{IT1} simultaneously.

Other  weak axial exchange and pion absorption amplitudes can be
derived from the operator amplitudes, constructed in
Ref.\cite{KT1} in conjunction with the Bethe--Salpeter equation,
by sandwiching them between the Dirac spinors for the free
nucleons. The weak axial exchange amplitudes of the $\rho$ meson
range are as follows. Besides the nucleon Born amplitude,
$J^a_{5\mu,\,\rho}$, the only potential contact term is
${J}^a_{5\mu,\,c\,\rho}(\pi)$. The mesonic amplitudes belong to
the non--potential ones and they are
${J}^a_{5\mu,\,a_1\,\rho}(a_1)$ and
${J}^a_{5\mu,\,a_1\,\rho}(\pi)$. Considered together with the
$\rho$ meson exchange amplitudes, the $a_1$ meson ones contain
only the potential amplitudes, that are the nucleon Born terms
$J^a_{5\mu,\,a_1}$ and three contact terms
${J}^a_{5\mu,\,c_i\,a_1}(\pi)$ (i=1,2,3). It can be verified that
separately, the exchange amplitudes ${ J}^a_{5\mu,\,\rho}(2)$ and
${ J}^a_{5\mu,\,a_1}(2)$, defined as \be
{J}^a_{5\mu,\,\rho}(2)\,=\,J^a_{5\mu,\rho}\,+\,{J}^a_{5\mu,\,c\,\rho}(\pi)\,+\,
{J}^a_{5\mu,\,a_1\,\rho}(a_1)\,+\,{J}^a_{5\mu,\,a_1\,\rho}(\pi)\,,
\label{JTREXBS} \ee and \be
{J}^a_{5\mu,\,a_1}(2)\,=\,J^a_{5\mu,\,a_1}\,+\,\sum\limits^3_{i=1}\,
{J}^a_{5\mu,\,c_i\,a_1}(\pi)\,, \label{JRA1EX} \ee respectively,
do not satisfy the standard PCAC constraint.

However, using a derivation analogous to that of Sect.\,3 of
Ref.\cite{KT1}, one finds \footnote{For more details see also
Sect.\,III of Ref.\,\cite{MRT}.} that the sum of the amplitudes
(\ref{JTREXBS}) and (\ref{JRA1EX}) satisfies the PCAC equation \be
q_\mu\left[\,{J}^a_{5\mu,\,\rho}(2)\,+\,{J}^a_{5\mu,\,a_1}(2)\,\right]\,=\,
i f_\pi m^2_\pi\,\Delta^\pi_F(q^2)\,\left[\, M^a_\rho(2)\, +\,
M^a_{a_1}(2)\,\right]\,.   \label{dJRPA1EX} \ee

The $\omega$ meson exchange amplitudes contain only the potential
amplitudes $J^a_{5\mu,\omega}$ and $J^a_{5\mu,\,c\,\omega}(\pi)$.
The divergence of the $\omega$ meson exchange amplitude
${J}^a_{5\mu,\,\omega}(2)$, defined as \be
{J}^a_{5\mu,\,\omega}(2)\,=\,J^a_{5\mu,\,\omega}\,+\,
J^a_{5\mu,\,c\,\omega}(\pi)\,, \label{Jaom} \ee yields the PCAC
equation \be q_\mu
J^a_{5\mu,\omega}(2)\,=\,if_\pi\,m^2_\pi\,\Delta^\pi_F(q^2)\,
M^a_\omega(2)\,.  \label{dpJacop} \ee

Using the results obtained in \cite{KT1}, one can derive the
two--nucleon weak axial and pion absorption amplitudes from the
Lagrangian \cite{STG} in the same manner. We mention only that the
$\rho$ and $a_1$ exchange amplitudes satisfy the PCAC separately
and that the model dependence should be expected at higher
energies, because it is of the short range nature. In the next
section, starting from the obtained two--nucleon weak axial
exchange amplitudes, we derive the WANECs of the heavy meson
range.

\section{Weak axial nuclear exchange currents  \label{CH2}}

We define the WANEC of the range B as \be
j^a_{5\mu,\,B}(2)\,=\,J^a_{5\mu,\,B}(2)\,-\,t^{a,\,FBI}_{5\mu,\,B}\,,
\label{JnaB} \ee where the two--nucleon amplitudes
$J^a_{5\mu,\,B}(2)$ are derived in the previous section, and
$t^{a,\,FBI}_{5\mu,\,B}$ is the first Born iteration of the
one--nucleon current contribution  to the two--nucleon scattering
amplitude, satisfying the Lippmann--Schwinger equation \cite{ATA},
\bea t^{a,\,FBI}_{5\mu,\,B}\,&=&\,V_B(\vec p^{\,\prime}_1,\vec
p^{\,\prime}_2;\vec P,\vec p_2)\frac{1}{P_0 - E(\vec P)+i\vep}
j^a_{5\mu}(1,\vec P,\vec p_1)\nonumber  \\
&&\,+\,j^a_{5\mu}(1,\vec p^{\,\prime}_1,\vec Q)\frac{1}{Q_0-E(\vec
Q)+i\vep}V_B(\vec Q,\vec p^{\,\prime}_2;\vec p_1,\vec p_2)
\,+\,\ot\,,  \label{LSE} \eea where $\vec P=\vec
p^{\,\prime}_1+\vec q_2=\vec p_1+\vec q$ and $\vec Q=\vec p_1-\vec
q_2=\vec p^{\,\prime}_1-\vec q$, as it is seen from
Fig.\,\ref{figg2}.  Further, $V_B$ is the one--boson exchange
nuclear potential and \be j^a_{5\mu}(1,\vec p^{\,\prime},\vec
p)\,=\,\bar u(p'){\hat J}^a_{5\mu}(1,1) u(p) \,, \label{onnc} \ee
is the weak axial nuclear one--nucleon current in the momentum
space. It is the non--relativistic reduction of the one--nucleon
amplitude $J^a_{\,5\mu}(1,1)$, given in Eq.\,({\ref{ONAC}). In the
nucleon kinematics, $\vec q_1=\vec p^{\,\prime}_1-\vec p_1$ and
$q_{10}=E(\vec p^{\,\prime}_1)-E(\vec p_1)$.
\begin{figure}[h!]
\centerline{ \epsfig{file=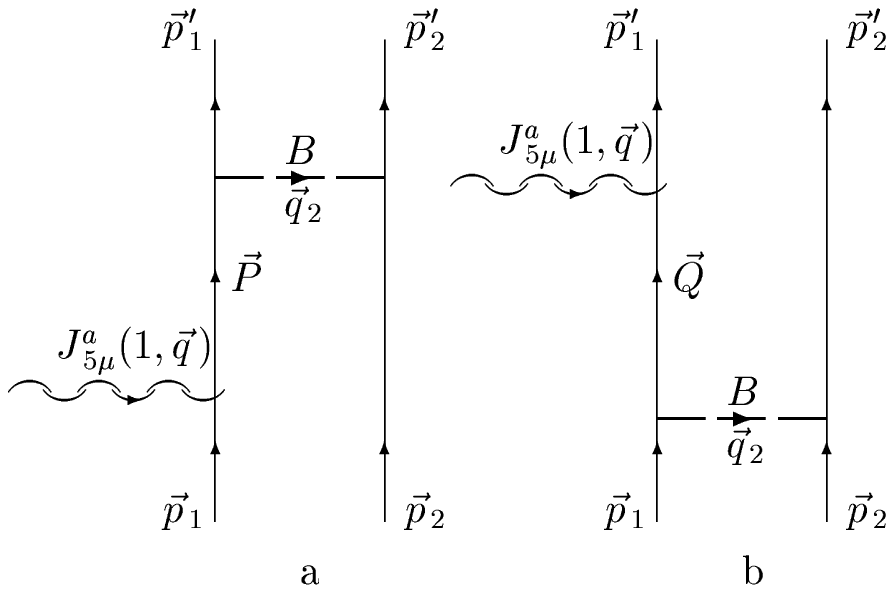} } \vskip 0.4cm \caption{The
kinematics of the first Born iteration. The nucleon line in the
intermediate state is on--shell. } \label{figg2}
\end{figure}
The amplitude $J^a_{5\mu,\,B}(2)$ contains the nucleon Born terms,
graphically presented in Figs.\,\ref{figg1}(a) and \ref{figg1}(b).
For them, however, the four--momentum conservation takes place.
Next we derive the continuity equation for the WANEC.

\subsection{The PCAC equation for the WANEC \label{CH21}}

In order to derive the nuclear PCAC for the current
$j^a_{5\mu,\,B}(2)$, we should know the continuity equation for
the first Born iteration $t^{a,\,FBI}_{5\mu,\,B}$. This equation
is \bea q_\mu t^{a,\,FBI}_{5\mu,\,B}\,&=&\,i f_\pi m^2_\pi
\Delta^\pi_F(\vec q^{\,2})\left[ V_B(\vec p^{\,\prime}_1,\vec
p^{\,\prime}_2;\vec P,\vec p_2)\frac{1}{P_0 - E(\vec P)+i\vep}
m^a(1,\vec P,\vec p_1) \right.\nonumber  \\
&& \left. +\,m^a(1,\vec p^{\,\prime}_1,\vec Q)\frac{1}{Q_0-E(\vec
Q)
+i\vep}V_B(\vec Q,\vec p^{\,\prime}_2;\vec p_1,\vec p_2)\right]\,-\,\left[V_B,\rho^a_5(1)\right] \nonumber  \\
&& +\,\ot\,,  \label{dtaFBI} \eea where \be
\left[V_B,\rho^a_5(1)\right]\,\equiv\,V_B(\vec p^{\,\prime}_1,\vec
p^{\,\prime}_2;\vec P,\vec p_2) \rho^a_5(1,\vec P,\vec
p_1)\,-\,\rho^a_5(1,\vec p^{\,\prime}_1,\vec Q) V_B(\vec Q,\vec
p^{\,\prime}_2;\vec p_1,\vec p_2)\,.  \label{cVrho} \ee It follows
from Eqs.\,(\ref{dJRPA1EX}) and (\ref{dpJacop}) that the
divergence of the amplitude $J^a_{5\mu,\,B}(2)$ can be written as
\be q_\mu J^a_{5\mu,\,B}(2)\,=\,i f_\pi m^2_\pi \Delta^\pi_F(\vec
q^{\,2})M^a_B(2)\,.  \label{dJaB} \ee Using Eqs.\, (\ref{dtaFBI})
and (\ref{dJaB}), we derive the PCAC for the WANEC, defined in
Eq.\,(\ref{JnaB}) \be q_\mu j^a_{5\mu,\,B}(2)\,=\,i f_\pi m^2_\pi
\Delta^\pi_F(\vec q^{\,2}){m}^a_B(2)\,
+\,\left(\left[V_B,\rho^a_5(1)\right]\,+\,\ot \right)\,,
\label{dWANEC} \ee where \be
{m}^a_B(2)\,=\,M^a_B(2)\,-\,m^{a,\,FBI}_B\,,  \label{bMaB} \ee and
the first Born iteration of the one--nucleon pion absorption
amplitude is \bea m^{a,\,FBI}_B\,&=&\,\,V_B(\vec
p^{\,\prime}_1,\vec p^{\,\prime}_2;\vec P,\vec p_2)
\frac{1}{P_0 - E(\vec P)+i\vep} m^a(1,\vec P,\vec p_1)  \nonumber  \\
&&  \,+\,m^a(1,\vec p^{\,\prime}_1,\vec Q)\frac{1}{Q_0-E(\vec Q)
+i\vep}V_B(\vec Q,\vec p^{\,\prime}_2;\vec p_1,\vec p_2)\,+\,\ot
\,. \label{MAFBIB} \eea We shall call the amplitude ${m}^a_B(2)$
the nuclear two--nucleon pion absorption amplitude of the range B.
It is seen from Eq.\,(\ref{bMaB}) that it is defined in the same
way as the WANEC in Eq.\,(\ref{JnaB}).

We now pass to investigate the structure of the WANECs.

\subsection{The structure of the WANECs \label{CH22}}

As it is seen from Eq.(\ref{JnaB}), the structure of the WANECs
differs from the structure of the two--nucleon weak axial exchange
amplitudes studied in the previous section. In the case of the
Schr\"odinger equation, the positive frequency part of the nucleon
Born term and the first Born iteration differ and this difference
provides a contribution to the WANECs. Here we shall calculate
this difference. For this purpose, we split the nucleon propagator
$S_F$ into the positive, $S_F^{(+)}$, and negative, $S_F^{(-)}$,
frequency parts (see App.\,\ref{happA}). Then the positive
frequency part $J^{a\,(+)}_{\,5\mu,\,B}$ of the nucleon Born term
$J^a_{\,5\mu,\,B}$, Eq.\,(\ref{JbBf}), can be cast into the form
\bea J^{a\,(+)}_{\,5\mu,\,B}&=&{\cal
V}_B(p'_1,p'_2;P,p_2)\frac{1}{P_0-E(\vec P)}{\tilde
J}_{\,5\mu}(1,P,p_1)
\frac{1}{2}(a^+ - a^-)  \nonumber  \\
&& +{\tilde J}_{\,5\mu}(1,p'_1,Q)\frac{1}{Q_0-E(\vec Q)}{\cal
V}_B(Q,p'_2;p_1,p_2) \frac{1}{2}(a^+ + a^-)+\ot\,,  \label{nBtpfp}
\eea where the quasipotential ${\cal V}_B$ is defined as \be {\cal
V}_B(p'_1,p'_2;p_1,p_2)\,=\,g^2_{BNN}{\bar u}(p'_1){\tilde {\cal
O}}^B_{(\lambda)}(-q_2)u(p_1)\, \Delta^B_{(\lambda
\eta)}(q_2)\,{\bar u}(p'_2){\tilde {\cal
O}}^B_{(\eta)}(q_2)u(p_2)\,,  \label{QPB} \ee and ${\tilde
J}_{\,5\mu}$ is the amplitude Eq.\,(\ref{ONAC}) without
$\tau^a/2$. For the amplitude $t^{a,\,FBI}_{5\mu,\,B}$,
Eq.\,(\ref{LSE}), one can write in a similar way \bea
t^{a,\,FBI}_{5\mu,\,B}\,&=&\,{\tilde V}_B(\vec p^{\,\prime}_1,\vec
p^{\,\prime}_2;\vec P,\vec p_2) \frac{1}{P_0 - E(\vec P)+i\vep}
{\tilde j}_{5\mu}(1,\vec P,\vec p_1)\frac{1}{2}(a^+ - a^-)\nonumber  \\
&&\,+\,{\tilde j}_{5\mu}(1,\vec p^{\,\prime}_1,\vec
Q)\frac{1}{Q_0-E(\vec Q)+i\vep} {\tilde V}_B(\vec Q,\vec
p^{\,\prime}_2;\vec p_1,\vec p_2)\frac{1}{2}(a^+ + a^-) +\ot\,,
\label{LSEB1} \eea where ${\tilde V}_B$ and ${\tilde j}_{5\mu}$
correspond to ${ V}_B$ and ${ j}^a_{5\mu}$, respectively, but
without the isospin dependence.

Using the definition Eq.\,(\ref{JnaB}) and Eqs.\,(\ref{nBtpfp})
and (\ref{LSEB1}), we can calculate the contribution to the WANEC
arising from the difference of the positive frequency nucleon Born
terms and the first Born iteration \bea
\Delta j^{a\,(+)}_{\,5\mu,\,B}\,&=&\,J^{a\,(+)}_{\,5\mu,\,B}\,-\,t^{a,\,FBI}_{5\mu,\,B} \nonumber  \\
\,&=&\,\frac{1}{P_0 - E(\vec
P)+i\vep}D_{\,\mu,\,B}(P)\frac{1}{2}(a^+ - a^-)
+\frac{1}{Q_0 - E(\vec Q)+i\vep}D_{\,\mu,\,B}(Q)\, \frac{1}{2}(a^+ + a^-) \nonumber  \\
&& \,+\,\ot\,,    \label{DJaB}  \\
D_{\,\mu,\,B}(P)&=&{\cal V}_B(p'_1,p'_2;P,p_2){\tilde
J}_{\,5\mu}(1,P,p_1) -{\tilde V}_B(\vec p^{\,\prime}_1,\vec
p^{\,\prime}_2;\vec P,\vec p_2)
{\tilde j}_{5\mu}(1,\vec P,\vec p_1)\,,  \label{DP}  \\
D_{\,\mu,\,B}(Q)&=&{\tilde J}_{\,5\mu}(1,p'_1,Q){\cal
V}_B(Q,p'_2;p_1,p_2)-{\tilde j}_{5\mu}(1,\vec p^{\,\prime}_1,\vec
Q) {\tilde V}_B(\vec Q,\vec p^{\,\prime}_2;\vec p_1,\vec p_2)\,.
\label{DQ} \eea Checking the structure of the currents and
potentials, one can see that the contributions to $\Delta
j^{a\,(+)}_{\,5\mu,\,B}$ can arise from the difference in the
dependence on the energy transfer of the BNN vertices, of the
currents, and of the B--meson propagators. We shall call them the
vertex, external and retardation currents, respectively. They
satisfy the equation,
 \be \Delta
j^{a\,(+)}_{\,5\mu,\,B}\,=\,j^{a}_{\,5\mu,\,B}(vert)\,
+\,j^{a}_{\,5\mu,\,B}(ext)\,+\,j^{a}_{\,5\mu,\,B}(ret)\,.
\label{SJs} \ee Besides these corrections, one obtains the whole
set of terms by the standard non--relativistic reduction of the
amplitudes $J^a_{\,5\mu,\,B}(2)$ up to the desired order in $1/M$.

\subsection{Results \label{CH23}}

In this section, we present the resulting WANECs.

\subsubsection{Vertex currents \label{CH231}}

As it is seen from Eqs.\,(\ref{Orho}) and (\ref{Oomega}), the
$\rho$-- and $\omega$ exchanges can contribute into the vertex
current $j^{a}_{\,5\mu,\,B}(vert)$. This contribution can be
calculated by expanding in the BNN vertex of the pseudopotential
${\cal V}_B$ around $P_0=E(\vec P)$ and $Q_0=E(\vec Q)$. Then one
obtains \bea {\cal V}_B(p'_1,p'_2;P,p_2)\,&=&\, {\tilde V}_B(\vec
p^{\,\prime}_1,\vec p^{\,\prime}_2;\vec P,\vec p_2)\,
+i g^2_{BNN}\frac{\kappa^B}{2M}[P_0-E(\vec P)]  \nonumber  \\
&& \times\,{\bar u}(p'_1)\sigma_{j4}u(P)\Delta^B_{j\eta}(q_2)
{\bar u}(p'_2)\tilde {\cal O}^{B,st}_\eta(q_2)u(p_2)\,,
\label{expVBP0} \eea and
\bea {\cal V}_B(Q,p'_2;p_1,p_2)\,&=&\,
{\tilde V}_B(\vec Q,\vec p^{\,\prime}_2;\vec p_1,\vec p_2)\,
+i g^2_{BNN}\frac{\kappa^B}{2M}[E(\vec Q)-Q_0]  \nonumber  \\
&& \times\,{\bar u}(Q)\sigma_{j4}u(p_1)\Delta^B_{j\eta}(q_2) {\bar
u}(p'_2)\tilde {\cal O}^{B,st}_\eta(q_2)u(p_2)\,.
\label{expVBQ0} \eea Substituting the expansions
Eqs.\,(\ref{expVBP0}) and (\ref{expVBQ0}) into
Eqs.\,(\ref{DJaB})--(\ref{DQ}) we obtain \bea
j^{a}_{\,5\mu,\,B}(vert)\,&=&\,i
g^2_{BNN}\frac{\kappa^B}{2M}\left[{\bar u}(p'_1)\sigma_{j4}u(P)
{\tilde j}_{5\mu}(1,\vec P,\vec p_1)\frac{1}{2}(a^+ - a^-)-
{\tilde j}_{5\mu}(1,\vec p^{\,\prime}_1,\vec Q)
{\bar u}(Q)\sigma_{j4}u(p_1) \right. \nonumber \\
&& \left.\,\times \frac{1}{2}(a^+ + a^-)\right]
\Delta^B_{j\eta}(q_2) {\bar u}(p'_2)\tilde {\cal
O}^{B,st}_\eta(q_2)u(p_2)\,+\,\ot\,. \label{D1JaB} \eea Performing
the non--relativistic reduction and preserving only the part
$\sim\,(1+\kappa^B)$, the space part of Eq.\,(\ref{D1JaB}) reduces
to \bea \vec {j}^{\,\,a}_{\,5,\,B}(vert)\,&=&\,
\frac{g^2_{BNN}}{(2M)^3}\kappa^B (1+\kappa^B) \bigg<g_A F_A\left\{
a^+ \left[\vec q \times \s2q2 +i\,\vec \sigma_1 \times (\vec P_1
\times \s2q2) \right]
 \right. \bigg. \nonumber \\
&&\left. \bigg. +i\, a^-\left[i \vec P_1 \times \s2q2 - \vec
\sigma_1 \times (\vec q \times \s2q2)\right]\right\}
\bigg. \nonumber \\
&& \bigg. + \frac{g_P}{m_l}\frac{\vec q}{2M} \bigg\{ia^+
\left[\vec q \cdot \vec \sigma_1\times(\vec P_1\times\s2q2)\right]
+ i a^-\bigg[i\,\vec q \ \cdot \vec P_1 \times \s2q2 \bigg. \bigg. \bigg.  \nonumber \\
&& \bigg. \bigg. \bigg. -\vec q \cdot \vec \sigma_1 \times (\vec q
\times \s2q2) \bigg] \bigg\} \bigg> \Delta^B_F(\vec q^{\,\,2}_2)
\,+\,\ot\,, \label{RDVECJPB} \eea where $\vec P_i=\vec
p^{\,\,\prime}_i+\vec p_i$, $g_P(\vec q^{\,\,2})/m_l=2g f_\pi
\Delta^\pi_F(\vec q^{\,\,2})$, and $m_l$ is the lepton mass. This
current is important for the $\rho$ meson exchange, since
$\kappa^B=\kappa^V_\rho\approx 6.6$.

Analogous calculations for the time component
$j^{a}_{\,50,\,B}(vert)$ show that in comparison with the space
component, Eq.\,(\ref{RDVECJPB}), it is by one order in 1/M
suppressed, \be j^{a}_{\,50,\,B}(vert)\,\approx\,{\cal
O}(1/M^4)\,. \label{TCDVECJPB} \ee According to the definition of
the exchange currents in Refs.\,\cite{KRT,TR}, only the negative
frequency part of the nucleon Born terms contributes. So the
vertex currents are absent in \cite{TR}. It means that the
currents \cite{TR} are suitable for calculations with the nuclear
wave functions that are solution of the equation of motion,
providing the first Born iteration that cancels exactly the
positive frequency part of the nucleon Born currents, in order to
avoid the double counting. In the pion exchange current the
analogous vertex current of our approach is again of the nominal
order ${\cal O}(M^{-3})$ \cite{AHHST,MRT1}, if the chiral model
with the pseudovector $\pi NN$ coupling is used. In the chiral
model with the pseudoscalar $\pi NN$ coupling, the vertex current
is absent, but the sum of the negative frequency part of the
nucleon Born term and of the PCAC constraint term yields the same
result \cite{ISTPR,TK1,MRT1}. Without the vertex current included,
the pion exchange current \cite{TR} is $\approx\,{\cal
O}(M^{-5})$.

\subsubsection{External exchange currents \label{CH232}}

These currents arise from the $q_0$ dependence of the amplitudes
$\tilde J_{5\,\mu}$ in Eqs.\,(\ref{DP}) and (\ref{DQ}). One
obtains \cite{MRT} \bea \vec
j^a_{\,5,\,B}(ext)\,&\approx&\,g_A\frac{\vq}{4M}\,\Delta^{a_1}_F(q^2)\,\left\{\,-
a^+[{\tilde V}^{(1)}_B(\vec q_2)
(\sa\cdot(\Pa+\q2))\,-\,(\sa\cdot(\Pa-\q2)){\tilde V}^{(1)}_B(\vec q_2)] \right. \nonumber \\
&& \left. + a^-[{\tilde V}^{(1)}_B(\vec
q_2)(\sa\cdot(\Pa+\q2))\,+\,(\sa\cdot(\Pa-\q2)){\tilde
V}^{(1)}_B(\vec q_2)]\, \right \}
\,+\,\ot\,,  \label{SjaextB}  \\
j^a_{50,\,B}(ext)\,&\approx&\,\frac{g_A}{2}\,[\Delta^{a_1}_F(q^2)-\Delta^{\pi}_F(q^2)]\,\left
\{\,a^+[{\tilde V}^{(1)}_B(\vec q_2)
(\sa\cdot\vq)\,-\,(\sa\cdot\vq){\tilde V}^{(1)}_B(\vec q_2)] \right. \nonumber \\
&& \left. - a^-[{\tilde V}^{(1)}_B(\vec
q_2)(\sa\cdot\vq)\,+\,(\sa\cdot\vq)){\tilde V}^{(1)}_B(\vec
q_2)]\, \right \} \,+\,\ot\,.  \label{TjaextB} \eea Here ${\tilde
V}^{(1)}_B$ is the leading order term in the B--meson exchange
potential ${\tilde V}_B$. In deriving these equations, we
neglected terms $\sim\,[\Delta^{B}_F(q^2)]^2$.

\subsubsection{Pair term retardation \label{CH233}}

This contribution arises from the pair terms due to the all meson
exchanges and it appears from the difference between the energy
dependence in the propagator of the pseudopotential and the
potential \cite{ATA}. The final result for the retardation
currents is \cite{MRT} \bea
j^{a}_{\,5\mu,\,B}(ret)\,&=&\,\frac{1}{4M}\left<\{(1+\nu)(\vec P_2
\cdot\vec q_2) +(1-\nu)[(\vec P_1+\vec q) \cdot\vec q_2)]\}{\tilde
V}^{(1)}_B(\vec q_2)
{\tilde j}_{5\mu}(1,\vec P,\vec p_1) \right.  \nonumber \\
&&\left.\,\times\,\frac{1}{2}(a^+ - a^-)\,-\,\{(1+\nu)(\vec P_2
\cdot\vec q_2)+(1-\nu)[(\vec P_1-\vec q) \cdot\vec q_2)]\}
{\tilde j}_{5\mu}(1,\vec p^{\,\prime}_1,\vec Q)\right. \nonumber \\
&&\left.\,\times\,{\tilde V}^{(1)}_B(\vec q_2)\frac{1}{2}(a^+ +
a^-)\right>\Delta^B_F(\vec q^{\,2}_2)\,+\,\ot\,. \label{JaretB}
\eea Here $\nu$ is a parameter of a unitary transformation
(cf.\,\cite{Fr}).

\subsubsection{Negative frequency part of the nucleon Born terms \label{CH234}}

This contribution arises from the nucleon Born currents
$J^{a}_{\,5\mu,\,B}$, Eq.\,(\ref{JbBf}), by the change
$S_F\,\rightarrow\,S^{(-)}_F$. We consider first the B=$\rho$ and
$\omega$ meson exchanges.

After the non-relativistic reduction, we obtain for the negative
frequency terms of the $\rho$-- and $\omega$ ranges \bea \vec
j^{\,\,a(-)}_{\,5,\,B}\,&=&\,\frac{g^2_{BNN}}{(2M)^3}(1+\kappa^B)\left<
\,a^+\left\{ g_A F_A\left[ (\vec \sigma_1 \cdot \vec q_2)\,\vec q
+ \vec q_2\times((\vec \sigma_1+\vec \sigma_2)\times\vec q_2)
\right.\right.\right.\nonumber \\
&&\left.\left.\left.\,+ i \vec P_1 \times (\vec \sigma_1 \times
(\vec \sigma_2\times \vec q_2)) - i(\vec P_1\times \vec
q_2)\right]
\,-\,2M\frac{g_P(q^2)}{m_l}(\vec \sigma_1 \cdot \vec q_2)\,\vec q\right\} \right. \nonumber \\
&&\left. \,-a^-\left\{g_A F_A\left[i(\vec \sigma_1\cdot \vec
\sigma_2 \times \vec q_2)\vec q +\vec P_1\times(\vec \sigma_2
\times \vec q_2) +i \vec q_2\times(\vec \sigma_1\times(\vec
\sigma_2 \times \vec q_2))
\right.\right.\right. \nonumber \\
&&\left.\left.\left.\,+\vec P_1\times(\vec \sigma_1\times \vec
q_2)\right] \,-\,2i\,M\frac{g_P(q^2)}{m_l}(\vec \sigma_1\cdot \vec
\sigma_2 \times \vec q_2)
\vec q \right\}\right>\,\Delta^B_F(\vec q_2^{\,\,2})\, \nonumber \\
&&\,+\,\frac{g^2_{BNN}}{(2M)^3}\left<\,a^+\,g_A
F_A\,\left[\,-i(\qb\times\Pb) \,+\,\Pa\times(\sa\times\Pb)\right]
\,-a^-\,\left\{g_A F_A\left[\,(\sa \cdot \Pb)\,\vq
\right.\right. \right.\nonumber \\
&& \left.\left.\left.
\,-\,i(\Pa\times\Pb)\,+\,\qb\times(\sa\times\Pb)\,\right]
\,-\,2M\frac{g_P(q^2)}{m_l}(\sa \cdot \Pb)\,\vq\right\} \right>
\,\Delta^B_F(\vec q_2^{\,\,2})+\,\ot\,, \label{JamBs} \\
j^{\,a(-)}_{\,50,\,B}\,&=&\,\frac{g^2_{BNN}}{(2M)^2}\,\left<\,(1+\kappa^B)\left\{-a^+
\left[ g_A F_A i (\vec \sigma_1\cdot \vec \sigma_2 \times \vec
q_2) + \frac{g_P(q^2)}{m_l}\,q_0\,(\vec \sigma_1 \cdot \vec
q_2)\right]
\right.\right. \nonumber \\
&&\left.\left. \,+ a^- \left[g_A F_A (\vec \sigma_1 \cdot \vec
q_2) +i  \frac{g_P(q^2)}{m_l}\,q_0\, (\vec \sigma_1\cdot \vec
\sigma_2 \times \vec q_2)\right]\right\}
\,-a^+\,g_A F_A\,(\sa\cdot\Pb) \right.  \nonumber \\
&&\left.\,+\,a^-\,\,\frac{g_P(q^2)}{m_l}\,q_0\,(\sa \cdot
\Pb)\,\right>\, \Delta^B_F(\vec q_2^{\,\,2}) \,+\,\ot\,.
\label{JamBt} \eea Comparing our current of Eq.\,(\ref{JamBs})
with the current ${\vec A}_\pm(V)$ of Ref.\,\cite{TR} we can see a
difference in the term of the form $\sa \qb^{\,\,2}$. Spurious
factor 1/4 appears in \cite{TR} due to the use of
$S^{(-)}_F(p)=(i\vec \gamma \cdot \vec p -M +\gamma_4 E(\vec
p^{\,\,2}))/4M^2$, and approximating $E=M$, when calculating the
contribution from the $\gamma_4 \gamma_4$ part of the vector
exchange. The same is true also for the scalar--isoscalar
exchange. The correct result is obtained by making use of
$E=M+{\vec p}^{\,\,2}/2M$.

Summing up the currents $\vec {j}^{\,\,a}_{\,5,\,B}(vert)$,
Eq.\,(\ref{RDVECJPB}), and $\vec j^{\,\,a(-)}_{\,5,\,B}$,
Eq.\,(\ref{JamBs}) for $B=\rho$, and keeping only the terms
$\sim\,(1+\kappa^V_\rho)$, we arrive at the nuclear potential term
of the $\rho$ range\footnote{Here we follow the nomenclature of
Ref.\,\cite{MRT1}.} \bea \vec
j^{\,\,a}_{\,5,\,\rho}(pot)\,&=&\,\left(\frac{g_\rho}{2}\right)^2
\frac{(1+\kappa^V_\rho)^2}{(2M)^3}\,g_A F_A\, \left\{\tau^a_2
\left[\vec q\times(\vec \sigma_2\times \vec q_2) + i\vec
\sigma_1\times (\vec P_1 \times
(\vec \sigma_2\times \vec q_2))\right] \right. \nonumber  \\
&& \left. \,+(\vec \tau_1\times\vec \tau_2)^a \left[i\vec
P_1\times (\vec \sigma_2\times \vec q_2) -\vec \sigma_1\times(\vec
q\times
(\vec \sigma_2\times \vec q_2))\right]\right\}\Delta^\rho_F(\vec q^{\,\,2}_2) \nonumber \\
&&\,-\left(\frac{g_\rho}{2}\right)^2\frac{(1+\kappa^V_\rho)}{(2M)^2}\,
\frac{g_P({\vec q}^{\,\,2})}{m_l}\,\left[\tau^a_2(\vec \sigma_1
\cdot \vec q_2) -(\vec \tau_1\times\vec \tau_2)^a  (\vec
\sigma_1\cdot \vec \sigma_2 \times \vec q_2)\right]
\vec q\,\Delta^\rho_F(\vec q^{\,\,2}_2) \nonumber \\
&&\,+\,\ot\,, \label{Japrs} \\
j^{\,a}_{\,50,\,\rho}(pot)\,&=&\,-\left(\frac{g_\rho}{2}\right)^2
\frac{(1+\kappa^V_\rho)}{(2M)^2}\big\{i\,g_A F_A \left[ \tau^a_2\,
(\vec \sigma_1\cdot \vec \sigma_2 \times \vec q_2) + (\vec
\tau_1\times\vec \tau_2)^a\,
(\vec \sigma_1 \cdot \vec q_2)\right]  \big. \nonumber \\
&&\big. \,+ \frac{g_P(q^2)}{m_l}q_0\left[\tau^a_2\,(\vec \sigma_1
\cdot \vec q_2) -  (\vec \tau_1\times\vec \tau_2)^a\,
 (\vec \sigma_1\cdot \vec \sigma_2 \times \vec q_2)\right]\big\}\,
\Delta^\rho_F(\vec q^{\,\,2}_2) \nonumber \\
&&\,+\,\ot\,.  \label{Japrt} \eea The part of our current
$j^{\,\,a}_{\,50,\,\rho}(pot)$, Eq.\,(\ref{Japrt}), that is
$\sim$\,g$_A$, is in agreement with the current
$J^{(2)a}_{50}(v-pair)$, Eq.\,(38a), derived by Towner \cite{To1}.
On the other hand, the part of our current  $\vec
j^{\,\,a}_{\,5,\,\rho}(pot)$, Eq.\,(\ref{Japrs}), that is
$\sim$\,g$_A$, differs from the $\rho$ meson pair term, used by
Schiavilla {\em et al.} \cite{Sch14} and also in
Ref.\,\cite{NSAPMGK}.

Since $\kappa_S\,\approx\,-0.12$, we keep only the negative
frequency Born term contribution into the omega meson potential
term, \bea \vec
j^{\,\,a}_{\,5,\,\omega}(pair)\,&=&\,\left(\frac{g_\omega}{2}\right)^2
\frac{1}{(2M)^3}\,\ta\,\left<\,g_A
F_A\left\{\,\Pa\times(\sa\times\Pb)\, -i(\qb\times\Pb)
+\,(1+\kappa_S)\,\big[\,\vq\,(\sa\cdot\qb) \right. \right. \big. \nonumber \\
&& \left.\left.\big.
\,-i(\Pa\times\qb)\,+\,\qb\times((\sa+\sb)\times\qb)
\,+\,i\Pa\times(\sa\times(\sb\times\qb))\,\big]\,\right\}\right.
 \nonumber \\
&&\left. \,-\frac{2M
g_P}{m_l}(1+\kappa_S)\,\vq\,(\sa\cdot\qb)\,\right>
\,\Delta^\omega_F(\vec q^{\,\,2}_2)\,+\,\ot\,.  \label{Japos} \\
j^{\,a}_{\,50,\,\omega}(pair)\,&=&\,-\left(\frac{g_\omega}{2}\right)^2
\frac{1}{(2M)^2}\,\ta\,\left\{\,g_A
F_A\big[(\sa\cdot\Pb)+i(1+\kappa_S)
(\sa\cdot\sb\times\qb)\big] \right. \nonumber \\
&&\left.
\,+\frac{g_P}{m_l}\,q_0\,(1+\kappa_S)(\sa\cdot\qb)\right\}
\,\Delta^\omega_F(\vec q^{\,\,2}_2)\,+\,\ot\,.  \label{Japot} \eea
In this case, the potential and pair terms coincide. Analogous
calculations for the a$_1$ meson exchange yield the pair term that
can be found in Ref.\,\cite{MRT}. The currents studied so far are
derived from the nucleon Born term amplitudes. We now present
shortly the  currents that have the origin in other amplitudes of
the section \ref{CH12}.

\subsubsection{Potential contact WANECs \label{CH235}}

Here we have the currents $j^a_{5\mu,\,c\rho}(\pi)$,
$j^a_{5\mu,\,c_i a_1}(\pi)$, i=1,2,3, and
$j^a_{5\mu,\,c\,\omega}(\pi)$, obtained by the non--relativistic
reduction of the pion production amplitudes, entering the weak
amplitudes $J^a_{5\mu,\,c\rho}(\pi)$, $J^a_{5\mu,\,c_i a_1}(\pi)$,
i=1,2,3, and $J^a_{5\mu,\,c\omega}(\pi)$, respectively. The
explicit results for the currents can be found in \cite{MRT}.

\subsubsection{Non--potential WANECs \label{CH236}}

In the considered model, there are two non--potential currents,
$j^a_{5\mu,\,a_1\rho}(a_1)$ and $j^a_{5\mu,\,a_1\rho}(\pi)$,
related to the mesonic amplitudes $J^a_{5\mu,\,a_1\rho}(a_1)$ and
$J^a_{5\mu,\,a_1\rho}(\pi)$, respectively \cite{MRT}. They are of
the $\rho$ meson range. In the next section, we shall use  the
space component of the current $j^a_{5\mu,\,a_1\rho}(a_1)$ in the
numerical estimates of the cross sections. It reads \bea {\vec
j}^{\,\,a}_{5,\,a_1\,\rho}(a_1)\,&=&\,\frac{1+\kappa^V_\rho}{2M}
\left(\frac{g_\rho}{2}\right)^2g_A F_A(q^2) (\vec \tau_1 \times
\vec \tau_2)^a\,\Delta^{a_1}_F(\qa^{\,\,2})
\left[ 2(\sb\cdot \qa\times \q2)\sa \right. \nonumber \\
&& \left. \,+ (\sa\cdot \vq) (\sb\times \q2) + (\q2\cdot \sa
\times \sb)\qa \right] \,\Delta^{\rho}_F(\q2^{\,\,2})\, +\,\ot\,.
\label{sjaa1ra1} \eea

We have completed the construction of the heavy meson WANECs of
our models. In the next section, we apply the leading terms of our
currents in calculations of the cross sections for the weak
deuteron disintegration by the low energy (anti)neutrinos.

\section{Numerical results    \label{CH3}}

We now calculate the cross sections for the reactions of the weak
deuteron disintegration by low energy (anti)neutrinos, \bea
\nu_x\,+\,d\,&\longrightarrow&\,\nu'_x\,+\,n\,+\,p\,, \label{NUD} \\
\overline{\nu}_x\,+\,d\,&\longrightarrow&\,\overline{\nu}'_x\,+\,n\,+\,p\,, \label{NCA} \\
\nu_e\,+\,d\,&\longrightarrow&\,e^-\,+\,p\,+\,p\,,  \label{CCN} \\
\overline{\nu}_e\,+\,d\,&\longrightarrow&\,e^+\,+\,n\,+\,n\,,
\label{CCA} \eea where $\nu_x$ $\bar{\nu}_x$ refers to any active
flavor of the (anti)neutrino. The reactions (\ref{NUD}) and
(\ref{CCN}) are important for studying the solar neutrino
oscillations, whereas the reactions (\ref{NCA}) and (\ref{CCA})
occur in experiments with reactor antineutrino beams. The precise
knowledge of the cross sections of  the $\nu d$ reactions is
needed \cite{AR} for the calculations of the response functions of
the SNO  detector \cite{SNO1,SNO2,SNO3,SNO4}. The total active
($\nu_x$) $^{8}Be$ solar neutrino flux \mbox{5.21 $\pm$ 0.27
(stat.) $\pm$ 0.38 (syst.) $\times$ 10$^6$ cm$^2$ s$^{-1}$} was
found in agreement with the standard solar models prediction.

Theoretical studies of the reactions (\ref{NUD})-(\ref{CCA}),
including exchange currents, were accomplished in the SNPA in
Refs.\,\cite{NSAPMGK,TKK,NSGK}, and in the EFT's in
Refs.\,\cite{ASPFK,MB1,MB2}. The calculations \cite{NSGK,YHH},
accomplished within the SNPA,  generally differ between themselves
at the threshold energies  by 5\%-10\% \cite{MB2}. In our opinion,
this provides a good motivation to make independent calculations.
Before presenting our numerical results, we introduce necessary
formalism and input data.

\subsection{Formalism \label{CH31}}

Making use of the technique, developed in Refs.\,\cite{JDW,ODW},
one obtains the following generic equation for the  cross sections
\bea \sigma\,&=&\,\frac{1}{6\pi^2}M_r G^2_W\, \int^{+1}_{-1} dx\,
\int^{k^{\,\prime}_{max}}_0\,dk^{\,\prime}\, \kappa_0
k^{\,\prime\,2} \,\left\{\sum_{\lambda j_f,\,J\ge
0}\left[\,j_0\,j^*_0|<\lambda j_f||{\hat M}_J||d>|^2
\right.\right.\nonumber \\ && \left.\left.\, +j_3\,j^*_3 |<\lambda
j_f||{\hat L}_J||d>|^2\, +\,2\Re \left(j_0 j^*_3\,<\lambda
j_f||{\hat M}_J||d>
<\lambda j_f||{\hat L}_J||d>^*\right) \right] \right. \nonumber \\
&&\left. +\,\fot({\vec j}\cdot {\vec j}^*\,-\,j_3
j^*_3)\sum_{\lambda j_f,\,J\ge 1}\left[ |<\lambda j_f||{\hat
T}^{\,mag}_J||d>|^2 \,+\,|<\lambda j_f||{\hat
T}^{\,el}_J||d>|^2\right] \right. \nonumber \\
&& \left. +\,i({\vec j}\times {\vec j}^*)_3\,\sum_{\lambda
j_f,\,J\ge 1}\Im \left( <\lambda j_f||{\hat
T}^{\,mag}_J||d><\lambda j_f||{\hat
T}^{\,el}_J||d>^*\right)\,\right\}\,. \label{CRS} \eea Here
$G_W=G_F$ for the neutral channel reactions (\ref{NUD}),
(\ref{NCA}), whereas $G_W=G_F\cos\theta_C$ for the charged channel
reactions (\ref{CCN}), (\ref{CCA}), $G_F$ is the Fermi constant
and $\theta_C$ is the Cabibbo angle. Further, the relative
momentum of the nucleons in the final state is \be
\kappa_0\,=\,\sqrt{2M_r(k-E'_l-\Delta)-\frac{1}{4}|\vec q|^2}\,,
\label{KAPPA0} \ee where $M_r$ is the reduced mass of the
nucleons, ${\vec k}({\vec k}^{\,\prime})$
 is the momentum of the incoming (anti)neutrino (outgoing lepton),
$E'_l=\sqrt{m^2_l+({\vec k}^{\,\prime})^2}$ is the energy of the
final lepton. For the reactions (\ref{NUD}) and (\ref{NCA}), \be
\Delta \,=\, |\epsilon_d|\,=\,2.2245 MeV, \ee whereas for the
reaction (\ref{CCN}), \be \Delta\, =\,
M_p-M_n+|\epsilon_d|\,=\,0.9312 MeV, \ee and for the reaction
(\ref{CCA}), \be \Delta = M_n-M_p+|\epsilon_d|=3.5178 MeV. \ee The
4-momentum transfer is \be \vec q\,=\,\vec k\,-\,{\vec
k}^{\,\prime}\,,\quad q_0=k-E'_l\,. \label{FMT} \ee For the
neutral current reactions, Eqs.\,(\ref{NUD}) and (\ref{NCA}), \be
k'_{max}\,=\,k x\,-\,4M_r\,+\,\left[\,8M_r(2M_r-\Delta+k(1-x))
+k^2(x^2-1) \right]^{\frac{1}{2}}\,. \label{EPMAX1} \ee For the
charged channel reactions, Eqs.\,(\ref{CCN}) and (\ref{CCA}), the
upper limit of the final lepton momentum, $k'_{max}$, is provided
by the solution of the equation \be
({\vec k}-{\vec k'})^2\,+\,8M_r(E'_l+\Delta-k)\,=\,0\,. \label{PLMAX} \ee The
lepton form factors entering Eq.\,(\ref{CRS}) are given by the
following equations \bea \fot({\vec j}\cdot {\vec j}^*\,-\,j_3
j^*_3)\,&=&\,2[\,1\,-\,({\hat k}\cdot{\hat q}) ({\vec \beta} \cdot
{\hat q})\,]\,\rightarrow\,\frac{4\sin ^2\frac{\theta}{2}}{|\vec
q|^2}
[\,|\vec q|^2\,+\,2 k k'\cos^2\frac{\theta}{2}\,] \,, \label{LFF1} \\
j_3 j^*_3\,&=&\,2[\,1\,-\,({\vec \beta}\cdot{\hat k})\,+\,2({\hat
k}\cdot{\hat q}) (\vec \beta \cdot {\hat
q})\,]\,\rightarrow\,\frac{4 q_0^2 \cos^2\frac{\theta}{2}}
{|\vec q|^2}\,, \label{LFF2} \\
j_0\,j^*_0\,&=&\,2[\,1\,+\,({\vec \beta}\cdot{\hat
k})\,]\,\rightarrow\,4\cos^2\frac{\theta}{2}
\,, \label{LFF3}  \\
j_0 j^*_3\,&=&\,2 {\hat q}\cdot({\hat k}\,+\,{\vec
\beta})\,\rightarrow\,
\frac{4 q_0}{|\vec q|}\,\cos^2\frac{\theta}{2}\,, \label{LFF4}  \\
i({\vec j}\times {\vec j}^*)_3\,&=&\,4 s {\hat q}\cdot({\hat
k}\,-\,{\vec \beta})\,\rightarrow\, \frac{8s}{|\vec
q|}\sin\frac{\theta}{2}\left[\,q^2\cos^2\frac{\theta}{2}\,+\,|\vec
q|^2 \sin^2\frac{\theta}{2}\,\right]^\fot\,. \label{LFF5} \eea The
form factors, presented after the arrows, are valid in the zero
mass limit of the outgoing lepton. The vector $\vec \beta$ and the
unit vectors of the type $\hat  b$ are defined as \be \vec
\beta\,=\,\frac{{\vec k}^{\,\,\prime}}{E'_l}\,,\quad {\hat
b}\,=\,\frac{\vec b}{|\vec b|}\,. \label{VB} \ee In addition, in
Eq.\,(\ref{LFF5}), $s=-1(+1)$ for the reactions (\ref{NUD}) and
(\ref{CCN}) ((\ref{NCA}) and (\ref{CCA})). The form factors
(\ref{LFF1})-(\ref{LFF4}) are the same for all four studied
reactions.

Let us note that  for the reaction (\ref{CCN}), the function under
integral in Eq.\,(\ref{CRS}) should be multiplied   by the Fermi
function. For the  solar neutrinos, a good approximation for this
function is \cite{MB2} \be
F(Z,E)\,=\,\frac{2\pi\nu}{1-\exp(-2\pi\nu)}\,,\quad
\nu=\alpha\,Z\,\frac{E'}{k'}\,. \label{FEFU} \ee The reduced
matrix elements in Eq.\,(\ref{CRS}) read \be <\lambda j_f||{\hat
O}_J||d>\,=\,\frac{\sqrt{4\pi}}{\kappa_0}\,\sum_{l',l_d}\,
\int_{0}^{\infty}\,r^2dr\,\frac{u^{j_f}_{l's,\,\lambda}(\kappa_0;r)}{r}\,
<(l's)j_f||{\hat O}_J||(l_d1)1>\,\frac{u_{l_d}}{r}\,,  \label{RME}
\ee where the nucleon-nucleon partial waves have asymptotics \be
u^{j}_{ls,\,\lambda}\,\rightarrow\,U^{j}_{ls,\,\lambda}\,\sin(\kappa_0
r \,-\,\frac{l\pi}{2} \,+\,\delta^j_\lambda)\,,  \label{ASYM} \ee
and the phase shifts and mixing parameters correspond to the
Blatt--Biedenharn \cite{BB} convention.

The multipoles are defined as \bea {\hat
T}^{\,mag}_{lm}(q)\,&=&\,\frac{(-i)^l}{4\pi}\int\,d\Omega_{\hat
q}\,
{\vec Y}^l_{lm}({\hat q})\,\cdot\,\vec j(\vq)\,,   \\
{\hat
T}^{\,el}_{lm}(q)\,&=&\,\frac{(-i)^{l-1}}{4\pi}\int\,d\Omega_{\hat
q}\,
{\vec Y}^{(1)}_{lm}({\hat q})\,\cdot\,\vec j(\vq)\,,   \\
{\hat L}_{lm}(q)\,&=&\,-\frac{(-i)^l}{4\pi}\int\,d\Omega_{\hat
q}\,
{\vec Y}^{(-1)}_{lm}({\hat q})\,\cdot\,\vec j(\vq)\,,   \\
{\hat M}_{lm}\,&=&\,\frac{(-i)^l}{4\pi}\int\,d\Omega_{\hat q}\,
Y_{lm}({\hat q})\, j_{0}(\vq)\,, \eea where \bea {\vec
Y}^{(1)}_{lm}({\hat q})\,&=&\,\sqrt{\frac{l+1}{2l+1}}{\vec
Y}^{l-1}_{lm}({\hat q})
+\sqrt{\frac{l}{2l+1}}{\vec Y}^{l+1}_{lm}({\hat q})\,,  \\
{\vec Y}^{(-1)}_{lm}({\hat q})\,&=&\,\sqrt{\frac{l}{2l+1}}{\vec
Y}^{l-1}_{lm}({\hat q})
-\sqrt{\frac{l+1}{2l+1}}{\vec Y}^{l+1}_{lm}({\hat q})\,.  \\
\eea The weak hadron neutral current, triggering the reactions
(\ref{NUD}) and (\ref{NCA}), is given by the equation \be
j_{NC,\,\mu}\,=\,(1-2 sin^2\theta_W)\,j^3_\mu-2 sin^2\theta_W\,
j_{S\mu}+j^3_{5\mu}\,,  \label{WHNC} \ee where $\theta_W$ is the
Weinberg angle \cite{DGH} ($sin^2\theta_W=0.23149\,(15)$
\cite{PDG}), $j^3_\mu$ ($j^3_{5\mu}$) is the third component of
the weak vector (axial) current in the isospin space, and
$j_{S\mu}$ is the isoscalar vector current. In its turn, the weak
hadron charged current is \be
j^a_{CC,\,\mu}\,=\,j^a_\mu\,+\,j^a_{5\mu}\,.  \label{WHCC} \ee At
low energies, the space component of the weak axial hadron current
is the most important one.

Our hadron currents consist of the one-- and two--nucleon parts.
The one--nucleon currents are of the form, \bea {\vec
j}^a\,&=&\,\frac{1}{2M}\,[F^V_1\,\Pa\,+\,iG^V_M
(\vec \sigma\times\vq)]\,\frac{\tau^a}{2}\,,  \label{ONCV}\\
{\vec j}^a_5\,&=&\,g_AF_A\{\vec \sigma\,-\,\frac{1}{8M^2}
[\Pa^{\,\,2}{\vec \sigma}\,-\,(\vec \sigma\cdot\Pa)\Pa\,+\,(\vec
\sigma\cdot\vq)\vq -i(\Pa\times\vq)]\}\frac{\tau^a}{2}\,.
\label{ONCA} \eea As to the two--nucleon part, we consider the
WANECs only. In addition to the new potential exchange currents
$\vec j^{\,\,a}_{\,5,\,\rho}(pot)$, $\vec
j^{\,\,a}_{\,5,\,\omega}(pair)$, and the non--potential exchange
current ${\vec j}^{\,\,a}_{5,\,a_1\,\rho}(a_1)$, we include in our
calculations the following exchange currents, derived in the
chiral invariant models \cite{TK1,CT,MRT1}:
\begin{enumerate}
\item The $\pi$ potential term, \bea \Delta\vec
j^{\,\,a}_{5,\,\pi}(pv)\,&=&\,\large(\frac{g}{2M}\large)^2\,\frac{g_A}{2M}\,
F_A(q^2)\, \large[\,(\vq + i \sa \times \Pa) \tba \,+\,(\Pa + i
\sa \times \vq) \t1t2a
\large\,]  \nonumber \\
&&\,\times \Delta^\pi_F(\qb^{\,\,2})\,F^2_{\pi NN}(\qb^{\,\,2})\,
(\sb \cdot \qb)\,+\,\ot\,.  \label{PIPT} \eea \item The
$\rho$-$\pi$ current, \bea \vec
j^{\,\,a}_{5,\,\pi}(\rho\pi)\,&=&\,-(\frac{g}{2M})^2\,\frac{1}{4Mg_A}\,
\large[\,1+\frac{m^2_\rho}{m^2_\rho+\qa^{\,\,2}}\,\large]\,\large[\,\Pa+
(1+\kappa^V_\rho)\,i\,(\sa \times \qa)\,\large] \nonumber \\
&&\,\times \,F_{\rho
NN}(\qa^{\,\,2})\,\Delta^\pi_F(\qb^{\,\,2})\,F_{\pi
NN}(\qb^{\,\,2})\, \,(\sb \cdot \qb)\t1t2a\,+\,\ot\,.
\label{RHPIT} \eea \item The $\Delta$ excitation current of the
pion range, \bea \vec
j^{\,\,a}_{5,\,\pi}(\Delta)\,&=&\,\frac{g_A}{9(M_\Delta-M)}\,F_A(q^2)\,
(\frac{f_{\pi N \Delta}}
{m_\pi})^2\,[\,4\qb\tba\,+\,i\,(\sa\times \qb)\t1t2a\,]  \nonumber  \\
&&\,\times \Delta^\pi_F(\qb^{\,\,2})\,F^2_{\pi NN}(\qb^{\,\,2})\,
(\sb \cdot \qb)\,+\,\ot\,.  \label{DETPR} \eea \item The $\Delta$
excitation current of the $\rho$ meson range, \bea \vec
j^{\,\,a}_{5,\,\rho}(\Delta)\,&=&\,-\frac{G_1 g^2_\rho
f_\pi}{9(M_\Delta-M)}\, \frac{f_{\pi N
\Delta}}{m_\pi}\,\frac{1+\kappa^V_\rho}{2M^2}\,
[\,4\qb\times(\sb\times \qb)\tba\,+\,i\,\sa\times(\qb\times
(\sb\times \qb))\t1t2a\,]  \nonumber  \\
&&\,\times \Delta^\rho_F(\qb^{\,\,2})\,F^2_{\rho
NN}(\qb^{\,\,2})\, +\,\ot\,.  \label{DETRR} \eea
\end{enumerate}
The form factors $F^V_1$, $G^V_M$, and $F_A$, are chosen in accord
with Ref.\,\cite{NSAPMGK}. The deuteron wave function and the wave
functions of the $^{1}S_0$ and $^{3}P_J$ states of the final
nucleons are obtained by solving the Schr\"odinger equation,
making use of the following first and second
generation realistic potentials: \\
(i) The potential OBEPQB \cite{Mac}, extended to include the $a_1$
meson exchange \cite{OPT}. The form factors, entering the BNN
vertices, are of the monopole (dipole) shape
for the exchanged meson B=$\pi$($\rho,\,\omega,\,a_1$).  \\
(ii) The Nijmegen 93 (Nijm93) and Nijmegen I (NijmI) potentials
\cite{SKTS}. In these potentials, the exponential strong BNN form
factors enter.

In order to keep our calculations consistent, we make use of the
same BNN form factors also in the currents. The value of the $\pi
N\Delta$ coupling, $f^2_{\pi N \Delta}/4\pi m^2_\pi$=0.7827
fm$^{-2}$, is derived from the $\Delta$ isobar width and it is
compatible with the pion photo- and electroproduction on  the
nucleon, the value of the constant $G_1$, $G_1$=2.525, follows
from the pion photo- and electroproduction \cite{DMW2}, and
$\kappa^V_\rho$=6.6 \cite{Mac}.

In table 1, we present the scattering lengths and the effective
ranges, obtained from the NijmI, Nijm93, OBEPQG and
AV18\footnote{This potential is used in Ref.\,\cite{NSGK}.}
\cite{AV18}
 potentials, and also the values of these quantities, applied in the EFT
calculations \cite{MB2}. For the generation of the final state
nucleon--nucleon wave functions from the NijmI and Nijm 93
potentials, we adopted the program COCHASE \cite{HLS}. This
program solves the Schr\"odinger equation by employing the
fourth--order Runge--Kutta method. This  can provide the
low--energy scattering parameters  slightly different from those,
obtained by the Nijmegen group, which makes use of the modified
Numerov method \cite{MRPC}. Some refit was necessary, in order to
get the required low--energy scattering parameters in the
neutron--proton and neutron--neutron $^{1}S_0$ states.
\par
Table 1. Scattering lengths and effective ranges (in fm) for the
nucleon--nucleon system in the $^{1}S_0$ state, corresponding to
the NijmI, Nijm93 \cite{SKTS}, OBEPQG \cite{OPT}, AV18 \cite{AV18}
potentials and as used in the EFT calculations \cite{MB2}, and the
experimental values.
\begin{center}
\begin{tabular}{|l||c|c|c|c||c||c|}\hline %\hline
         & NijmI & Nijm93   & OBEPQG &  AV18 &  EFT  &        exp.    \\\hline\hline
$a_{np}$ & -23.72 & -23.74  & -23.74 &-23.73 &-23.7  &   -23.740$\pm$0.020$^1$  \\
$r_{np}$ &   2.65 &   2.68  &   2.73 &  2.70 &  2.70 &     2.77
$\pm$0.05$^1$ \\\hline
$a_{pp}$ &  -7.80 &  -7.79  &  -     & -7.82 & -7.82 &    -7.8063$\pm$0.0026$^2$ \\
$r_{pp}$ &   2.74 &   2.71  &  -     &  2.79 &  2.79 &
2.794$\pm$0.014$^2$  \\\hline
$a_{nn}$ & -18.16 & -18.11  & -18.10 &-18.49 &-18.5  &    -18.59$\pm$0.40$^3$   \\
$r_{nn}$ &   2.80 &   2.78  &   2.77 &  2.84 &  2.80 &
2.80$\pm$0.11$^4$   \\\hline
\end{tabular}\\
$^1$ Ref.\,\cite{CDB};\, $^2$ Ref.\,\cite{BCSS};\, $^3$
Ref.\,\cite{MSL};\, $^4$ Ref.\,\cite{TG}
\end{center}

Below, we compare  our numerical results for the cross sections
with those obtained in the pionless EFT \cite{MB2} and in the
models, developed within the SNPA \cite{NSAPMGK,NSGK,YHH}.
However, in order to make the comparison with the
Ref.\,\cite{MB2}, one needs to know the constant $L_{1,\,A}$,
entering the cross sections of the pionless EFT. In the next
section, we extract it from the numerical values of the cross
sections, obtained in various potential models.

\subsection{Extraction of the low energy constant $L_{1,\,A}$ \label{CH32}}

In Ref.\,\cite{MB2}, the effective cross sections for the
reactions (\ref{NUD})-(\ref{CCA}) are  presented in the form \be
\sigma_{EFT}(E_\nu)\,=\,a(E_\nu)\,+\,L_{1,\,A}\,b(E_\nu)\,.
\label{sigB} \ee The amplitudes $a(E_\nu)$ and $b(E_\nu)$ are
tabulated in \cite{MB2} for each of the reactions
(\ref{NUD})--(\ref{CCA}), from the lowest possible  (anti)neutrino
energy up to 20 MeV. The constant $L_{1,\,A}$ cannot be determined
from reactions between elementary particles. In our analysis
\cite{MRTDEB}, we extracted $L_{1,\,A}$ from our cross sections,
calculated in the same approximation, as it was used in
Ref.\,\cite{MB2}: only the $^{1}S_0$ wave was taken into account
in the nucleon--nucleon final states and the nucleon variables
were treated non-relativistically. The knowledge of $L_{1,\,A}$
allowed us to compare our cross sections with
$\sigma_{EFT}(E_\nu)$. The results showed that the EFT cross
sections were in better agreement with the cross sections of
Ref.\,\cite{NSGK}, even though Nakamura {\em et al.} took into
account also the contribution from the $^{3}P_J$ waves and treated
the phase space relativistically. We defined the extracted value
of the constant $L_{1,\,A}$ as an average value,
$\bar{L}_{1,\,A}$, according to the equation \be
\bar{L}_{1,\,A}\,=\,\frac{\sum^N_{i=1}\,L_{1,\,A}(i)}{N}\,,\quad
L_{1,\,A}(i)\,=\,\frac{\sigma_{pot,i}\,-\,a_i}{b_i}\,, \label{LB}
\ee where $\sigma_{pot,i}$ is the cross section, calculated in the
potential model and for the {\it i-th} (anti)neutrino energy. The
application of the same equation in the present calculations
yields the values of $\bar{L}_{1,\,A}$, presented in table 2. As
it is seen  from  table 2, the variation of  $\bar{L}_{1,\,A}$ is
about 10 \% for the neutral current reactions, whereas it is up to
40 \% for the charged current reactions.

Let us note that the values of the constant $L_{1,\,A}$, obtained
from various analyses \cite{BY2,MB3}, are charged with large
errors. E.g.\,, the analysis \cite{MB3} of the data provided by
the experiments with the reactor antineutrino beams yielded
$\bar{L}_{1,\,A}=3.6\pm 5.5$ $fm^3$.
\par
Table 2. Values of the constant  $\bar{L}_{1,\,A}$ (in $fm^3$),
extracted from the cross sections of the reactions
(\ref{NUD})-(\ref{CCA}), which were calculated with the NijmI,
Nijm93,  and OBEPQG potentials, and from the cross sections NSGK
taken from table I of Ref.\cite{NSGK}.
\begin{center}
\begin{tabular}{|l||c|c|c||c|}\hline %\hline
reaction&   NijmI & Nijm93 & OBEPQG & NSGK   \\\hline\hline
(\ref{NUD}) & 4.8 & 5.4 & 5.0 &   5.4   \\\hline (\ref{NCA}) & 5.2
& 5.8 & 5.4 &   5.5 \\\hline (\ref{CCN}) & 4.4 & 5.3 & - & 6.0
\\\hline (\ref{CCA}) & 4.8 & 5.7 & 7.2 &  5.6   \\\hline
\end{tabular}
\end{center}
In the next section, employing the extracted values of the
constant $\bar{L}_{1,\,A}$, we compare  our cross sections with
those of the pionless EFT \cite{MB2}, and also with the cross
sections of Refs.\,\cite{NSAPMGK,NSGK,YHH}.

\subsection{Comparison of the cross sections \label{CH33}}

For the comparison presented in table 3 and table 4, we employ the
cross sections calculated with the nuclear wave functions
generated from the NijmI potential. In comparing our results with
those of Ref.\,\cite{MB2} we make use of the values of the weak
interaction constants $G_F=1.166\times 10^{-5} GeV^{-2}$ and
$g_A=-1.26$, but we adopt the value $g_A=-1.254$ when comparing
with Refs.\, \cite{NSGK,YHH}. In the calculations of the cross
sections for the charged channel reactions (\ref{CCN}) and
(\ref{CCA}), the value of the Cabibbo angle is taken $\cos
\theta_C = 0.975$.
\par
Table 3. Cross sections and the differences in \% between the
cross sections for the reactions (\ref{NUD}) and (\ref{NCA}). The
first 7 columns is related to the reaction (\ref{NUD}). In the
first column, $E_\nu$ [MeV] is the neutrino energy, in the second
column, $\sigma_{NijmI}$ (in $10^{-42}\times$ cm$^2$) is the cross
section,  calculated with the NijmI nuclear wave functions. Column
3 reports the difference between $\sigma_{Nijm I}$ (NijmI) and the
EFT cross section (\ref{sigB}) $\sigma_{EFT}$, calculated with the
corresponding constant $\bar{L}_{1,\,A}$ of table 2. The
difference between  $\sigma_{NSGK}$, taken from table I of
Ref.\,\cite{NSGK}, and $\sigma_{EFT}$, is reported in column 4.
Further, $\Delta_{1(2)}$ is the difference  between the cross
sections $\sigma_{NijmI}$ ($\sigma_{Nijm93}$) and
$\sigma_{NSGK}$; $\Delta_3$ is the difference between the cross
sections $\sigma_{NijmI}$ and $\sigma_{YHH}$, where the cross
section $\sigma_{YHH}$ is  taken from table I of Ref.\,\cite{YHH}.
The second part of the table is an analogue  for the reaction
(\ref{NCA}).
\begin{center}
\begin{tabular}{|l c c c c c c||l c c c c c c|}\hline
& & $ \nu_x$  +  $d$ & $\longrightarrow $ & ${\nu_x}'$ +  $n\,p$ &
& & & & ${\bar \nu}_x$  +  $d$ & $\longrightarrow $ & ${{\bar
\nu}_x}'$ +  $n\,p$ & &
\\\hline
$E_\nu$
&$\sigma_{NijmI}$&NijmI&NSGK&$\Delta_1$&$\Delta_2$&$\Delta_3$&
$E_{\bar{\nu}}$&$\sigma_{NijmI}$&NijmI&NSGK&$\Delta_1$&$\Delta_2$&$\Delta_3$
\\\hline\hline
3&0.00335&1.0&0.4&-1.1&-0.5&-& 3&0.00332&0.3&0.1&-1.1&-0.5&-
\\
4&0.0306 &1.0&0.2&-0.8&-0.2&12.0& 4&0.0302 &0.7&0.2&-0.8&-0.1&9.3
 \\
5&0.0948 &1.0&0.2&-0.9&-0.2&5.0& 5&0.0929 &0.7&0.1&-0.8&-0.1&0.9
 \\
6&0.201  &0.8&0.1&-1.0&-0.3&10.2& 6&0.196  &0.8&0.3&-0.8&-0.1&5.8
 \\
7&0.353  &0.8&0.1&-1.1&-0.3&8.2& 7&0.342  &0.5&0.1&-0.9&-0.2&2.0
 \\
8&0.552  &0.7&0.2&-1.2&-0.4&10.1& 8&0.532 &1.1&0.8&-1.0&-0.2&3.2
 \\
9&0.799  &0.8&0.4&-1.3&-0.5&9.0& 9&0.766 &0.6&0.2&-1.1&-0.3&1.0
 \\
10&1.095 &0.2&-0.1&-1.4&-0.6&7.8& 10&1.045&0.4&0.2&-1.2&-0.4&-1.5
 \\
11&1.441 &0.6&0.5&-1.6&-0.8&9.6& 11&1.368&-0.1&-0.2&-1.3&-0.5&-0.4
 \\
12&1.836&-0.2&-0.3&-1.8&-0.9&8.8&
12&1.734&-0.3&-0.4&-1.4&-0.5&-2.5
 \\
13&2.282&-0.1&0.0&-1.9&-1.1&10.2&
13&2.144&-0.3&-0.2&-1.5&-0.6&-1.7
\\
14&2.779&-0.3&0.0&-2.1&-1.2&9.9& 14&2.597&-0.4&-0.2&-1.6&-0.7&-3.4
 \\
15&3.326&-0.6&-0.1&-2.3&-1.4&10.8&
15&3.092&-0.5&-0.2&-1.8&-0.9&-3.5
 \\
16&3.923&-0.9&-0.3&-2.5&-1.6&10.5&
16&3.628&-0.6&-0.1&-1.9&-1.0&-4.9
\\
17&4.571&-1.2&-0.4&-2.7&-1.8&11.2&
17&4.206&-0.8&-0.2&-2.1&-1.1&-5.2
 \\
18&5.269&-1.4&-0.3&-3.0&-2.0&11.1&
18&4.824&-1.0&-0.3&-2.3&-1.3&-6.6
 \\
19&6.017&-1.7&-0.4&-3.2&-2.2&10.6&
19&5.481&-1.3&-0.3&-2.5&-1.5&-6.9
 \\
20&6.814&-2.1&-0.6&-3.5&-2.5&11.7&
20&6.177&-1.4&-0.2&-2.7&-1.6&-8.1
 \\\hline
\end{tabular}
\end{center}
Comparing this table with table 3 and table 4 of
Ref.\,\cite{MRTDEB} we conclude that the effect of the $^{3}P_J$
waves leads our cross sections in better agreement with the EFT
cross sections, but the effect is weaker in comparison with the
one obtained in Ref.\,\cite{NSGK}. On the other hand, the
disagreement with the cross sections of Ref.\,\cite{YHH} is even
more pronounced at higher energies. Table 3 also shows that our
cross sections are in better agreement with the other cross
sections in the case of the reaction (\ref{NCA}), than in the case
of the reaction (\ref{NUD}). One can also conclude from the
differences, given in the columns NijmI, NSGK, $\Delta_1$, and
$\Delta_2$ that the cross sections for the reactions (\ref{NUD})
and (\ref{NCA}) are described by both the potential models and the
pionless EFT with an accuracy better than 3\%.

In table 4, we present the comparison of the cross sections for
the  reactions in the charged channel.
\par
Table 4. Cross sections and the differences in \% between cross
sections for the reactions (\ref{CCN}) and (\ref{CCA}). For
notations, see table 3.
\begin{center}
\begin{tabular}{|l c c c c c c||l c c c c c c|}\hline
& & $ \nu_e$  +  $d$ & $\longrightarrow $ & $e^-$ +  $p\,p$ & & &
& & ${\bar \nu}_e$  +  $d$ & $\longrightarrow $ & $e^+$ +  $n\,n$
& &
\\\hline
$E_\nu$
&$\sigma_{NijmI}$&NijmI&NSGK&$\Delta_1$&$\Delta_2$&$\Delta_3$&
$E_{\bar{\nu}}$&$\sigma_{NijmI}$&NijmI&NSGK&$\Delta_1$&$\Delta_2$&$\Delta_3$
\\\hline\hline
2&0.00338&-5.8&-0.6&-7.6&-6.7&-&
-&-&-&-&-&-&- \\
3&0.0455 &-0.8&-0.3&-3.0&-2.0&-&
-&-&-&-&-&-&- \\
4&0.153  & 0.1&-0.6&-1.9&-0.9&1.9&
-&-&-&-&-&-&- \\
5&0.340  & 1.1& 0.1&-1.6&-0.6&2.9&
5&0.0274 &-1.7&-0.9 &-2.5&-1.6&9.0 \\
6&0.613  & 1.5& 0.4&-1.5&-0.5&3.1&
6&0.116  &-0.3 &-0.1 &-2.1&-1.1&8.1 \\
7&0.979  & 1.6& 0.4&-1.5&-0.5&3.1&
7&0.277  &-0.1 &-0.2 &-1.8&-0.7&7.4 \\
8&1.440  &-0.4&-2.4&-1.6&-0.6&3.3&
8&0.514  &0.2 &-0.1 &-1.6&-0.5&7.1 \\
9&2.000  &-0.6&-2.3&-1.7&-0.6&3.1&
9&0.830  &0.1 &-0.2 &-1.6&-0.5&7.0 \\
10&2.662 &-0.2&-1.7&-1.9&-0.7&3.3&
10&1.226 &0.7  &0.3 &-1.6&-0.4&6.9 \\
11&3.426 & 3.4& 3.3&-2.0&-0.9&3.1&
11&1.701 &0.6  &0.2 &-1.6&-0.5&6.2\\
12&4.294 & 1.0& 0.3&-2.2&-1.1&2.9&
12&2.255 &0.5  &0.1 &-1.6&-0.5&6.4 \\
13&5.268 & 0.8& 0.2&-2.4&-1.3&2.9&
13&2.887 &0.3  &0.0 &-1.7&-0.6&5.9\\
14&6.348 & 0.5& 0.2&-2.7&-1.5&2.7&
14&3.596 &0.4  &0.2 &-1.8&-0.6&5.6\\
15&7.535 & 0.3& 0.2&-2.9&-1.7&2.4&
15&4.380 &0.1  &0.0 &-2.0&-0.7&5.5\\
16&8.829 &-0.1&-0.1&-3.1&-2.0&2.2&
16&5.237& 0.1  &0.1 &-2.1&-0.9&5.3\\
17&10.23 &-0.5&-0.1&-3.5&-2.3&1.9&
17&6.167& 0.0  &0.2 &-2.3&-1.0&4.4 \\
18&11.74 &-0.8&-0.1&-3.8&-2.6&1.1&
18&7.168& 0.1  &0.4 &-2.4&-1.1&4.2 \\
19&13.36 &-1.0&-0.0&-4.1&-2.9&1.0&
19&8.237&-0.2  &0.2 &-2.6&-1.3&3.9 \\
20&15.09 &-1.5&-0.3&-4.4&-3.2&1.0& 20&9.373&-0.3  &0.3
&-2.8&-1.5&3.7  \\\hline
\end{tabular}
\end{center}
It can be seen from the left-hand part of table 4 that our cross
sections and the cross section \cite{NSGK} are smooth functions of
the neutrino energy, whereas the EFT cross section exhibits abrupt
changes in the region $7\,<\,E_\nu\,<\,12$ MeV. In our opinion,
the reason can be in an incorrect treatment of the Coulomb
interaction between protons in the EFT calculations. It follows
from the right-hand part of table 4 that our cross sections and
also  the cross sections by Nakamura {\em et al.} \cite{NSGK} are
in very good agreement with the EFT cross sections, in spite of
the fact that the difference up to 3 \% between these calculations
persists. Large difference between our calculations and those by
Ying {\em et al.} at the threshold energies can be understood by a
poor description of the neutron--neutron low energy scattering
parameters by the Paris potential model.

Next we compare our cross sections with those of
Ref.\,\cite{NSAPMGK}. For this purpose, we use $G_F=1.1803\times
10^{-5}$ GeV$^{-2}$ and \mbox{$g_A=-1.267$}. The results,
displayed in table 5 are for the reaction (\ref{NUD}) and the
OBEPQG potential. The parameters of this potential, required in
the calculations of the exchange current effects, are $g^2_{\pi
NN}/4\pi$=14.4, $g^2_{\rho NN}/4\pi$=3.6, $g^2_{\omega
NN}/4\pi$=98, $\Lambda_\pi$=8.62 fm$^{-1}$,
$\Lambda_\rho$=$\Lambda_\omega$=9.38 fm$^{-1}$, and
$\Lambda_{a_1}$=10.14 fm$^{-1}$. The shape of the table enables
one to compare  our calculations directly with those, presented in
table 1 and table 3 of Ref.\,\cite{NSAPMGK}. It follows from table
5 that the most important contributions to the cross section are
from the $\Delta$ excitation currents and from other currents of
the pion range. It is also seen that the contributions from the
$\pi$ potential- and the $\rho$-$\pi$ terms compensate each other
to a large extent \cite{MRT1}. One can also see that  the currents
interfere between themselves destructively. Besides, the
contributions from the heavy meson exchange currents, such as the
$\rho$ potential term, the $\omega$ pair- and $a_1$-$\rho$
currents, are at the threshold energies numerically insignificant.
In the last row, we present the cross sections due to the
transition $^{3}S_1-^{3}D_1\,\rightarrow\,^{1}S_0$ only. The
comparison of the last two rows shows that the effect of the
$^{3}P_J$ interaction in the final state is
$\approx\,\,0.2/0.6/1.2$ \% at $E_\nu=10/15/20\,\,MeV$.
\par
Table 5. Cumulative contributions to the cross section
$\sigma_{\nu d}$ ($\times 10^{-42}$ cm$^2$) from the weak axial
exchange currents  for various neutrino energies are displayed.
The cross section, calculated in the impulse approximation
(including the relativistic corrections of the order ${\cal
O}(1/M^2)$ is presented in the row labelled as IA ($\delta$IA).
Other contributions correspond to the exchange currents as
follows: $\Delta(\pi)\rightarrow\vec j^{\,\,3}_{5,\,\pi}(\Delta)$;
$\Delta(\rho)\rightarrow\vec j^{\,\,3}_{5,\,\rho}(\Delta)$;
$p(\pi)\rightarrow \Delta\vec j^{\,\,a}_{5,\,\pi}(pv)$;
$p(\rho)\rightarrow\vec j^{\,\,3}_{5,\,\rho}(pot)$;
$p(\omega)\rightarrow\vec j^{\,\,3}_{5,\,\omega}(pair)$;
$\rho$-$\pi\rightarrow\vec j^{\,\,3}_{5,\,\pi}(\rho\pi)$;
$a_1$-$\rho\rightarrow\vec j^{\,\,3}_{5,\,a_1\rho}(a_1)$. The
cross section in the n-th row is given by the contribution of all
previous currents, the n-th current including. The number in the
bracket is the ratio of the n-th cross  section to the cross
section in the row above. The first three exchange current
contributions are from the long--range exchange currents. The
cross sections in the last row are obtained with the
neutron--proton wave function calculated only for the transition
$^{3}S_1-^{3}D_1\,\rightarrow\,^{1}S_0$.
\begin{center}
\begin{tabular}{|l||c|c|c|c|}\hline %\hline
$E_\nu$ [MeV]& 5 & 10 & 15 & 20 \\\hline
IA & 0.0948 (-) & 1.088(-) & 3.296 (-) & 6.740 (-) \\
$+\delta$IA & 0.0944 (0.997) & 1.084 (0.996) & 3.281 (0.995) & 6.709 (0.995) \\
+$\Delta(\pi)$ & 0.0996 (1.055) & 1.151 (1.062) & 3.499 (1.067) & 7.175 (1.070) \\
+p($\pi$) & 0.0989 (0.992) & 1.141 (0.991) & 3.466 (0.991) & 7.010 (0.990)  \\
+$\rho$-$\pi$ & 0.0997 (1.008) & 1.152 (1.010) & 3.502 (1.010) & 7.181 (1.011)  \\
+$\Delta(\rho)$ & 0.0983 (0.986) & 1.134 (0.984) & 3.444 (0.983) & 7.056 (0.983)  \\
+p($\rho$) & 0.0986 (1.003) & 1.137 (1.003) & 3.455 (1.003) & 7.079 (1.003) \\
+p($\omega$) & 0.0986 (1.000) & 1.138 (1.001) & 3.456 (1.001) & 7.083 (1.001) \\
+$a_1$-$\rho$ & 0.0985 (0.999) & 1.136 (0.999) & 3.452 (0.999) &
7.074 (0.999) \\\hline\hline $\sigma(^{1}S_0)$   & 0.0984 & 1.134
& 3.432 & 6.993 \\\hline
\end{tabular}\end{center}

However, the currents, adopted in Ref.\,\cite{NSAPMGK} differ from
our currents in several aspects:
\begin{enumerate}
\item The $\pi$ pair term is constructed from the pseudoscalar
$\pi NN$ coupling. This makes the model incompatible with the
chiral invariance. \item The $\rho$-$\pi$ current, used in
\cite{NSAPMGK}, is connected to our current (\ref{RHPIT}) by the
change \be 1+\frac{m^2_\rho}{m^2_\rho+\qa^{\,\,2}}\,\rightarrow\,
\frac{2m^2_\rho}{m^2_\rho+\qa^{\,\,2}}\,. \label{CH} \ee \item The
$\Delta$ excitation currents (\ref{DETPR}) and (\ref{DETRR}) are
in \cite{NSAPMGK} supplied with the couplings of the constituent
quark model, additionally multiplied by a factor 0.8. This factor
was found to be needed to obtain the experimental value of the
Gamow--Teller matrix element for the triton $\beta$ decay
\cite{Sch14}. However, this way of fixing the MECs is a model
dependent procedure depending on the specific choice of the
nuclear model which is the AV18 potential in this case. Let us
note that the exchange current model \cite{NSAPMGK,Sch14}
underestimates \cite{MSRKV} the precise data on the ordinary muon
capture in $^{3}He$ , $\mu^- + ^{3}He\rightarrow \nu_\mu + ^{3}H$.
Besides, the resulting $\pi N \Delta$ coupling is in a sharp
contradiction with its value obtained  from the $\Delta$ width and
also from  the pion photo- and electroproduction on the nucleon.
On the other hand, the damping factor $M/M_\Delta \approx 0.8$
arises in the vector and axial $\Delta$ excitation currents, if
they are constructed from the 'gauge symmetric' Lagrangians
\cite{PAS,SMT}. It means that in order to be consistent, one
should repeat the analysis of the reaction
$n+p\,\rightarrow\,d+\gamma$ \cite{NSGK} with the damped vector
$\Delta$ excitation currents and make use of them also in
\cite{NSAPMGK}. Let us note that other possibility to reconciliate
the Gamow--Teller matrix element of the triton $\beta$ decay with
the data is to vary the cutoff parameters $\Lambda_\pi$ and
$\Lambda_\rho$ \cite{SWIS}, keeping the couplings of the $\Delta$
excitation  currents fixed. \item The structure of our $\rho$
potential term is different from the $\rho$ pair term
\cite{NSAPMGK}, and the $\omega$ pair term and the $a_1$-$\rho$
current are in \cite{NSAPMGK} absent. \item The BNN vertices are
of the monopole shape. \item As it has been discussed in
\cite{MRT1}, it is not clear which nuclear continuity equation the
currents \cite{NSAPMGK,Sch14} satisfy.
\end{enumerate}

In table 6, we present the results of calculations obtained with
our currents modified according to points 2, 3 and 5 above and
with the values of the parameters \cite{NSAPMGK} $g^2_{\pi
NN}/4\pi$=14.81, $g^2_{\rho NN}/4\pi$=2.0, $f^2_{\pi N
\Delta}/4\pi m^2_\pi$=0.4701 fm$^{-2}$, $\Lambda_\pi$=4.8
fm$^{-1}$, $\Lambda_\rho$=6.8 fm$^{-1}$.

In comparing our table 6 with table 3/Model I of
Ref.\,\cite{NSAPMGK}, we mention our stronger $\pi$ potential term
and weaker $\rho$ potential- and $\rho$-$\pi$ terms. Moreover, the
$\pi$- and $\rho$ potential terms are of the opposite sign in
comparison with the $\pi$- and $\rho$ pair terms of
Ref.\cite{NSAPMGK}.

The comparison of our $\rho$-$\pi$ row of table 6 with the rows
labelled as NPC Bonn and NPC AV18 shows that our results are by
1\% to 2\% smaller, while the same comparison in table 5 exhibits
an opposite effect.

Table 6. Cumulative contributions to the cross section
$\sigma_{\nu d}$ ($\times 10^{-42}$ cm$^2$) from the modified weak
axial exchange currents  for various neutrino energies. For the
notations, see table 5. The modification of the currents and the
choice of the parameters are described in the text. The values of
the cross sections in the rows labelled as NPC Bonn and NPC AV18
are displayed for comparison \cite{NPC}. These cross sections
correspond to the calculations \cite{NSAPMGK}, restricted to the
transition $^{3}S_1$-$^{3}D_1\,\rightarrow\,^{1}S_0$, and for the
CD-Bonn \cite{CDB}, and the AV18 \cite{AV18} potentials,
respectively.
\begin{center}
\begin{tabular}{|l||c|c|c|c|}\hline %\hline
$E_\nu$ [MeV]& 5 & 10 & 15 & 20  \\\hline
IA & 0.0948 (-) & 1.088(-) & 3.296 (-) & 6.741 (-) \\
$+\delta$IA & 0.0944 (0.997) & 1.084 (0.996) & 3.281 (0.995) & 6.709 (0.995)  \\
+$\Delta(\pi)$ & 0.0962 (1.019) & 1.107 (1.021) & 3.356 (1.023) & 6.869 (1.024) \\
+p($\pi$) & 0.0956 (0.994) & 1.099 (0.993) & 3.331 (0.992) & 6.814 (0.992) \\
+$\rho$-$\pi$ & 0.0957 (1.001) & 1.100 (1.001) & 3.334 (1.001) & 6.821 (1.001) \\
+$\Delta(\rho)$ & 0.0955 (0.997) & 1.097 (0.997) & 3.323 (0.997) & 6.798 (0.997) \\
+p($\rho$) & 0.0956 (1.001) & 1.099 (1.002) & 3.329 (1.002) & 6.810 (1.002) \\
+p($\omega$) & 0.0957 (1.001) & 1.099 (1.000) & 3.332 (1.001) & 6.816 (1.001) \\
+$a_1$-$\rho$ & 0.0956 (0.999) & 1.099 (1.000) & 3.330 (1.000) &
6.813 (1.000) \\\hline $\sigma(^{1}S_0)$   & 0.0956 & 1.097 &
3.311 & 6.732 \\\hline
NPC Bonn & 0.09552 & 1.099 & 3.323 & 6.765   \\
NPC AV18 & 0.09565 & 1.101 & 3.332 & 6.787   \\\hline
\end{tabular}\end{center}
Comparing table 6 with table 5 one concludes that in this model,
the exchange effect is much weaker. Now also the $\rho$-$\pi$
current is numerically insignificant.

Besides the calculations presented in table 5 and table 6, that
are obtained for the reaction (\ref{NUD}) with the OBEPQG wave
functions, we calculated also the  cumulative cross sections for
all the reactions (\ref{NUD})-(\ref{CCA}) using the wave functions
generated from the NijmI potential \cite{SKTS}. The results are
presented in tables 7, 8, 9 and 10. In this case, we chose the
couplings $G_F$, $g_A$ and $\cos\theta_C$ according to
Refs.\,\cite{EB},\cite{MS}, \be
G_F\,=\,1.16637(1)\,\times\,10^{-5}\,GeV^{-2}\,,\quad
g_A\,=\,-1.2720(8)\,,\quad \cos\theta_C\,=\,0.9730\pm 0.0004\pm
0.0012\pm 0.0002\,.  \label{WCO} \ee Here the Fermi constant $G_F$
is  fixed by the muon decay. Making use of this value of $G_F$,
the constants $g_A$ and $\cos\theta_C$ are extracted from the
neutron beta decay.

Table 7. Cumulative contributions to the cross section
$\sigma_{\nu d}$ ($\times 10^{-42}$ cm$^2$) for the reaction
(\ref{NUD}), calculated with the nuclear wave functions  and
couplings and cutoffs, entering the axial exchange current
operators, that correspond to the NijmI potential, for various
neutrino energies. For the notations, see table 3.
\begin{center}
\begin{tabular}{|l||c|c|c|c|}\hline %\hline
$E_\nu$ [MeV]& 5 & 10 & 15 & 20  \\\hline
IA & 0.0927 (-) & 1.065(-) & 3.227 (-) & 6.599 (-)\\
$+\delta$IA & 0.0923 (0.996) & 1.061 (0.996) & 3.212 (0.995) & 6.567 (0.995)  \\
+$\Delta(\pi)$ & 0.0977 (1.057) & 1.130 (1.065) & 3.436 (1.070) & 7.047 (1.073) \\
+p($\pi$) & 0.0969 (0.992) & 1.119 (0.991) & 3.401 (0.990) & 6.972 (0.989) \\
+$\rho$-$\pi$ & 0.0977 (1.009) & 1.131 (1.010) & 3.439 (1.011) & 7.051 (1.011) \\
+$\Delta(\rho)$ & 0.0962 (0.984) & 1.111 (0.982) & 3.374 (0.981) & 6.912 (0.980) \\
+p($\rho$) & 0.0965 (1.003) & 1.114 (1.003) & 3.384 (1.003) & 6.934 (1.003) \\
+p($\omega$) & 0.0966 (1.002) & 1.116 (1.002) & 3.392 (1.002) &
6.951 (1.002) \\\hline $\sigma(^{1}S_0)$   & 0.0966 & 1.114 (99.8)
& 3.372 (99.4) & 6.868 (98.8)\\\hline
\end{tabular}
\end{center}

Table 8. Cumulative contributions to the cross section
$\sigma_{\nu d}$ ($\times 10^{-42}$ cm$^2$) for the reaction
(\ref{NCA}), calculated with the nuclear wave functions  and
couplings and cutoffs, entering the axial exchange current
operators, that correspond to the NijmI potential, for various
antineutrino energies. For the notations, see table 3.
\begin{center}
\begin{tabular}{|l||c|c|c|c|}\hline %\hline
$E_{\bar\nu}$ [MeV]& 5 & 10 & 15 & 20  \\\hline\hline
IA & 0.0908 (-) & 1.016(-) & 2.996 (-) & 5.972 (-) \\
$+\delta$IA & 0.0905 (0.996) & 1.011 (0.996) & 2.982 (0.995) & 5.942 (0.995)  \\
+$\Delta(\pi)$ & 0.0957 (1.058) & 1.079 (1.067) & 3.198 (1.072) & 6.398 (1.077)  \\
+p($\pi$) & 0.0949 (0.992) & 1.069 (0.990) & 3.165 (0.990) & 6.327 (0.989) \\
+$\rho$-$\pi$ & 0.0958 (1.009) & 1.080 (1.011) & 3.201 (1.011) & 6.402 (1.012)\\
+$\Delta(\rho)$ & 0.0943 (0.984) & 1.060 (0.982) & 3.138 (0.980) & 6.270 (0.979) \\
+p($\rho$) & 0.0945 (1.003) & 1.064 (1.003) & 3.148 (1.003) & 6.291 (1.003)\\
+p($\omega$) & 0.0947 (1.002) & 1.066 (1.002) & 3.156 (1.002) &
6.307 (1.003) \\\hline $\sigma(^{1}S_0)$   & 0.0947 & 1.064 (99.8)
& 3.136 (99.4) & 6.225 (98.7) \\\hline
\end{tabular}
\end{center}

Table 9. Cumulative contributions to the cross section
$\sigma_{\nu d}$ ($\times 10^{-42}$ cm$^2$) for the reaction
(\ref{CCN}), calculated with the nuclear wave functions  and
couplings and cutoffs, entering the axial exchange current
operators, that correspond to the NijmI potential, for various
neutrino energies. For the notations, see table 3.
\begin{center}
\begin{tabular}{|l||c|c|c|c|}\hline %\hline
$E_\nu$ [MeV]& 5 & 10 & 15 & 20 \\\hline\hline
IA & 0.3320 (-) & 2.587(-) & 7.300 (-) & 14.59 (-)\\
$+\delta$IA & 0.3309 (0.997) & 2.577 (0.996) & 7.269 (0.996) & 14.52 (0.995) \\
+$\Delta(\pi)$ & 0.3481 (1.052) & 2.729 (1.059) & 7.730 (1.063) & 15.49 (1.066) \\
+p($\pi$) & 0.3458 (0.993) & 2.708 (0.992) & 7.666 (0.992) & 15.35 (0.991) \\
+$\rho$-$\pi$ & 0.3491 (1.010) & 2.737 (1.011) & 7.752 (1.011) & 15.53 (1.012) \\
+$\Delta(\rho)$ & 0.3438 (0.985) & 2.690 (0.983) & 7.611 (0.982) & 15.23 (0.981)  \\
+p($\rho$) & 0.3447 (1.003) & 2.698 (1.003) & 7.635 (1.003) & 15.28 (1.003) \\
+p($\omega$) & 0.3452 (1.001) & 2.703 (1.002) & 7.649 (1.002) &
15.31 (1.002) \\\hline $\sigma(^{1}S_0)$   & 0.3451 & 2.696 (99.7)
& 7.597 (99.3) & 15.12 (99.8) \\\hline
\end{tabular}
\end{center}

Table 10. Cumulative contributions to the cross section
$\sigma_{\nu d}$ ($\times 10^{-42}$ cm$^2$) for the reaction
(\ref{CCA}), calculated with the nuclear wave functions  and
couplings and cutoffs, entering the axial exchange current
operators, that correspond to the NijmI potential, for various
antineutrino energies. For the notations, see table 3.
\begin{center}
\begin{tabular}{|l||c|c|c|c|}\hline %\hline
$E_{\bar\nu}$ [MeV]& 5 & 10 & 15 & 20 \\\hline\hline
IA & 0.0268 (-) & 1.190(-) & 4.233 (-) & 9.031 (-) \\
$+\delta$IA & 0.0267 (0.997) & 1.185 (0.996) & 4.213 (0.995) & 8.985 (0.995)   \\
+$\Delta(\pi)$ & 0.0281 (1.051) & 1.259 (1.062) & 4.505 (1.069) & 9.653 (1.074) \\
+p($\pi$) & 0.0279 (0.993) & 1.249 (0.992) & 4.464 (0.991) & 9.559 (0.990) \\
+$\rho$-$\pi$ & 0.0281 (1.009) & 1.263 (1.011) & 4.518 (1.012) & 9.680 (1.013) \\
+$\Delta(\rho)$ & 0.0277 (0.985) & 1.240 (0.982) & 4.428 (0.980) & 9.475 (0.979) \\
+p($\rho$) & 0.0278 (1.002) & 1.244 (1.003) & 4.443 (1.003) & 9.509 (1.004) \\
+p($\omega$) & 0.0278 (1.000) & 1.246 (1.002) & 4.452 (1.002) &
9.529 (1.002) \\\hline $\sigma(^{1}S_0)$   & 0.0278 & 1.244 (99.8)
& 4.426 (99.4) & 9.407 (98.7) \\\hline
\end{tabular}
\end{center}
It is seen from table 7--table 10 that again, the effects of the
$\pi$ potential- and $\rho$-$\pi$ terms compensate each other
considerably. But in this potential model, the $\rho$ potential-
and $\omega$ pair terms interfere additively and the sum of them
contribute sensibly. The numbers in the brackets show the part (in
\%) of the $^{1}S_0$  cross section from the total cross section.

In table 11, we compare the $^{1}S_0$ phase shifts (in degrees)
for the  neutron--proton scattering obtained from different
potentials, used in the calculations of the cross sections,
presented above.

Table 11. Comparison  of the $^{1}S_0$ phase shift (in degrees)
for the  neutron--proton scattering described by the potentials
CD-Bonn \cite{CDB},
OBEPQG \cite{OPT}, NijmI and Nijm93 \cite{SKTS}.\\
\begin{center}
\begin{tabular}{|l||c|c|c|c|c|}\hline %\hline
E$_{lab}$ [MeV]& 1 & 5 & 10 & 25 & 50 \\\hline\hline
CD-Bonn & 62.09 & 63.67 & 60.01 & 50.93 & 40.45\\
OBEPQG & 62.02 & 63.56 & 59.91 & 50.82 & 40.13 \\
NijmI  & 62.12 & 63.74 & 60.10 & 51.04 & 40.56 \\
Nij93  & 62.05 & 63.61 & 59.94 & 50.85 & 40.38 \\\hline
\end{tabular}
\end{center}

The phase shifts labelled as CD-Bonn were delivered by R.
Machleidt \cite{MAC04}. As it is seen, the resulting phase shifts
are very close among themselves up to energies of 50 MeV. It means
that all four potentials are in the $^{1}S_0$ channel about the
same for the internucleon distances larger  than 0.6 fm.

%%%%%%%%%%%%%%%%%%%%%%%%%%%%%%%%%%%%%%%%%%%%%%%%%%%%%%%%%%%%%%%%%%%%%%%%%%%%%%

In Fig.\,\ref{figg3}, we consider for the reaction (\ref{NUD}) and
for the strongest multipole ${\hat T}^{el}_1$ the ratios of the
radial density for various exchange currents to the density of the
$\Delta$ excitation current of the pion range. As it is seen from
Eq.\,(\ref{CRS}), the densities depend on the cosine of the
scattering angle, $x=\cos\theta$, and on the energy $\nu'$ of the
outgoing neutrino. In its turn, $\nu'_{max}$ is given by the
incident neutrino energy $\nu$ and $x$.   As it is seen from the
figure, the ratios of the densities decrease substantially with
increasing values of $r$ for the short--range WANECs.

\begin{figure}[h!]
\centerline{ \epsfig{file=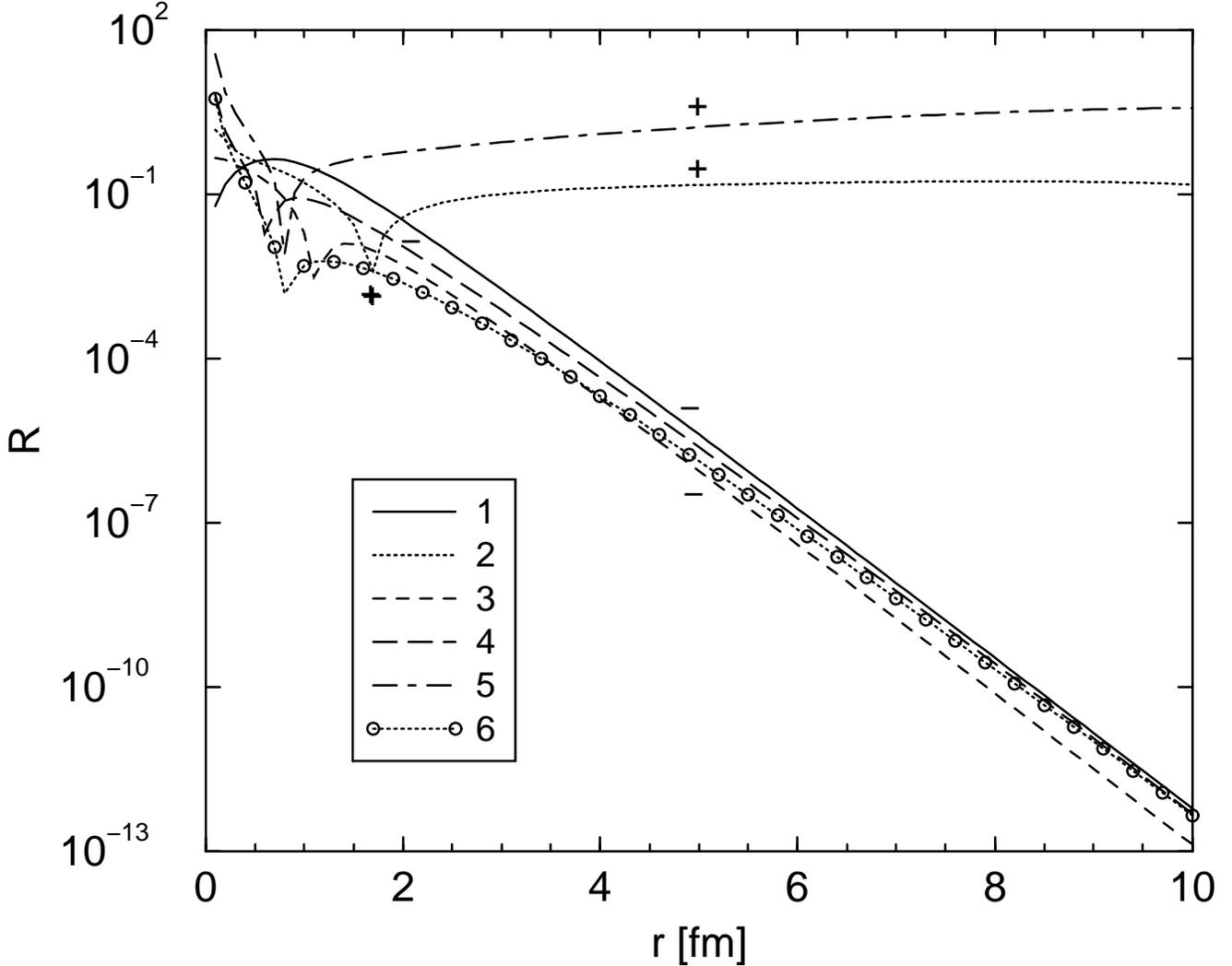} } \vskip 0.4cm \caption{The
ratios of the radial density of various WANECs to the density of
the $\Delta$ excitation current of the pion range, $\Delta(\pi)$,
are presented for the reaction (\ref{NUD}), and for the ${\hat
T}^{el}_1$ multipole. The selected kinematics is $\nu=5$ MeV,
$\theta=45^\circ$, $\nu'=\nu'_{max}/2$. With this choice, the
momentum transfer is $|\vec q|=4.12$ MeV. The curves correspond to
the following WANECs: 1- $\Delta(\rho)$, 2- p($\pi$), 3-
p($\rho$), 4- p($\omega$), 5- $\rho-\pi$, 6- $a_1-\rho$. The sign
+ (-) at a branch of a curve means that the ratio of the densities
is positive (negative). The signs of the branches before and/or
after the kinks are opposite. } \label{figg3}
\end{figure}

\section{Discussion of obtained results    \label{CH4}}

In this paper, we construct the weak axial nuclear heavy meson
exchange currents of the $\rho$--, $\omega$-- and a$_1$ meson
range in the TAA. These currents are suitable for calculations
employing the nuclear wave functions obtained by solving the
Schr\"odinger equation with the OBEPs. Including the $\pi$
exchange current, our model currents  provide the realistic space
component of the axial exchange currents. Adopting for the
exchange currents the needed input (couplings, cutoffs) from the
OBEPs, we can made a reliable estimate of the effect due to these
exchanges. Since our Lagrangians belong generically to the
non--linear realization of the chiral group, the $\sigma$ meson is
explicitly lacking. On the other hand, the OBEPs contain the
$\sigma$ meson exchange, needed to explain the medium range
attraction. Then it follows that the exchange effect, arising from
this meson, cannot be calculated consistently in our approach.
However, the structure of the space component of the WANECs of the
$\sigma$ range does not allow it to play any significant role at
low energies. This is discussed in more detail in appendix
\ref{happB}.

In accord with Ref.\,\cite{ATA}, we define the nuclear exchange
currents in Eq.\,(\ref{JnaB}) as the difference of the
relativistic Feynman amplitudes of Sect.\,\ref{CH1} and of the
first Born iteration of the nuclear equation of motion
(\ref{LSE}). The currents, derived in this manner, are given by a
finite set of terms and they satisfy the nuclear PCAC equation
(\ref{dWANEC}). Let us note that our approach follows methods
standardly used in the few--body nuclear physics at low and
intermediate energies: the transition operator that is constructed
for free nucleons is sandwiched between the nuclear wave functions
obtained by solving the Schr\"odinger equation with realistic
potentials. As is well known from the textbooks, if the transition
operator is the sum of the one- and two--nucleon vector currents,
it should satisfy the nuclear form of the CVC hypothesis
(\ref{NCVC}). Then accepting chiral symmetry as one of the
corner--stones on which the nuclear physics calculations rely, one
should construct the axial currents satisfying the nuclear PCAC
constraint, as it is done in this paper. If the particles cannot
be considered as free (nuclear matter) or non--relativistic then
one should use other methods to describe the nuclear states, and
to construct the currents in conjunction with the relevant
equation of motion for nuclear states, as it was done, e.g., in
Ref.\,\cite{KT1} for the axial currents and the Bethe--Salpeter
equation, or in Refs.\,\cite{CoR,JW} for the electromagnetic
currents and the Blankenbecler--Sugar--Logunov-Tavkhelidze
quasipotential equation.

After performing the standard non--relativistic reduction, we
obtain the currents in the leading order in 1/M. The non--zero
difference arises due to the difference between the positive
frequency part of the nucleon Born terms and of the first Born
iteration, which gives rise to the vertex, external and
retardation terms. The sum of the vertex terms and of the negative
frequency Born terms (pair terms) results in the nuclear potential
currents. The vertex currents contribute essentially in the case
of the $\pi$- and $\rho$ meson exchanges. Note that the space
parts of our $\pi$- and $\rho$ potential currents (\ref{Japrs})
differ from the the space parts of the $\pi$- and $\rho$ pair
terms, used in Refs.\,\cite{NSAPMGK,Sch14,TR}.

Besides the contribution from the nucleon Born terms, the
potential contact and non--potential currents appear after the
non--relativistic reduction. These currents have never been
constructed before. An infinite set of such terms was supposed to
exist in Ref.\,\cite{TR}, where it was argued that these terms
should be numerically insignificant because of the short--range
suppression present in nuclear wave functions. We have shown here
that the method based on the chiral invariance provides a finite
set of terms, of which contributions can be estimated
quantitatively.

The application of our currents to the reaction of  weak deuteron
disintegration by (anti)neutrinos at low energies shows that our
currents differ from the set that has recently been used in
Ref.\,\cite{NSAPMGK}, in that (i) The $\pi$- and $\rho$ potential
terms contribute with the opposite sign, in comparison with the
$\pi$- and $\rho$ pair terms \cite{NSAPMGK}, respectively. (ii)
The effect of the $\pi$ potential term cancels strongly the
effect, arising from the $\rho$-$\pi$ current. Besides we have
shown that at the threshold energies, the contributions to the
cross sections from the heavy meson exchanges interfere
destructively in conjunction with the OBEPQG potential, and the
heavy meson effect is strongly suppressed in comparison with the
effect arising from the currents of the pion range, as supposed in
\cite{TR}. However, in conjunction with the potential NijmI, the
$\rho$ potential- and $\omega$ pair terms interfere additively,
and this sum is non-negligible. This result is important in
particular in the case, if the $\Delta$ isobar currents are
adopted with the suppressed strength \cite{NSAPMGK}. Such analysis
can be performed for other potentials and/or energies, too, thus
having bookkeeping of the exchange current effects under control.
But at higher energies, other parts of our currents, not
considered in these calculations, should be also estimated.

We compared our cross sections with those obtained in the EFT
\cite{MB1,MB2} in terms of the constant $L_{1,\,A}$. Our values of
the constant  $L_{1,\,A}$ are somehow smaller than its value
obtained in the analysis of Refs.\,\cite{MB2,NSGK}, but in a good
agreement with the dimensional analysis and with the data analysis
\cite{BY2,MB3}. The comparison of the cross sections for the
reactions in the neutral current channel and for the reaction
(\ref{CCA}) shows that they are described by both the potentials
models and the pionless EFT with an accuracy better than 3 \%. For
the reaction (\ref{CCN}), the achieved accuracy is $\approx\,3.4$
\%. In this case, abrupt changes in the EFT cross section in the
interval $7\,<\,E_\nu\,<\,12\,MeV$ appeared, whereas  the cross
sections obtained from the potential models are smooth.

Our exchange charge and current densities are obtained in a
general reference frame in terms of individual nucleon
coordinates. This is a good approximation in the considered energy
region, since the estimated effect of the boost currents is
negligible. For higher energies, the calculations of the
observables by sandwiching the operators between the intrinsic
wave functions require to go one step further: by analogy with the
case of the electromagnetic exchange currents \cite{AA}, the
center--of--mass frame dependence should be separated. This will
be done elsewhere.

\section*{Acknowledgments}

This work is supported in part by the grants GA \v{C}R 202/03/0210
and 202/06/0746, and by Ministero dell' Istruzione, dell' Universit\`a e della
Ricerca of Italy (PRIN 2003). We thank R. Machleidt, M.
Rentmeester and S. Nakamura for the correspondence and J. Adam,
Jr. for the discussions.

\appendix

\section{Definitions and notations}
\label{happA}

We use the Pauli metrics. The free nucleon spinors are normalized
according to \be u^+(p)u(p)\,=\,v^+(p)v(p)\,=\,1\,.  \label{uvn}
\ee The spinors are \be u(p)\,=\,\sqrt{\frac{E(\vec p)+M}{2E(\vec
p)}} \left(\begin{array}{c} 1\\ \frac{\vec \sigma \cdot \vec
p}{E(\vec p)+M} \end{array} \right)\,,\quad
v(p)\,=\,\sqrt{\frac{E(\vec p)+M}{2E(\vec
p)}}\left(\begin{array}{c}
                      \frac{\vec \sigma \cdot \vec p}{E(\vec p)+M}
                      \\ 1 \end{array} \right)\,.  \label{uv}
\ee The nucleon propagator is split into the positive- and
negative frequency parts as follows \be S_F(p)\,=\,-\frac{1}{i\not
p + M}\,\equiv\,S_F^{(+)}(p)\,+\,S_F^{(-)}(p)\,,  \label{SF} \ee
where \be S_F^{(+)}(p)\,=\,\frac{1}{p_0 - E(\vec p)}u(p)\bar
u(p)\,,\quad S_F^{(-)}(p)\,=\,\frac{1}{p_0 + E(\vec p)}v(-p)\bar
v(-p)\,.    \label{SFPM} \ee For a weak semileptonic reaction in
the two--nucleon system NN, \be
NN(P_i)\,+\,l_i(p_i)\,\longrightarrow\,NN(P_f)\,+\,l_f(p_f)\,,
\label{spNN} \ee we write the field--theoretical S--matrix element
in the form \be
S\,=\,i(2\pi)^4\,\delta^{(4)}(P_f+p_f-P_i-p_i)\,\tilde
l_\mu(0)W^a_\mu(q)\,,  \label{Sm} \ee where the matrix element of
the lepton weak current is \be \tilde
l_\mu(0)\,=\,\left<l_f,p_f|l_\mu(0)|l_i,p_i\right>\,, \label{lme}
\ee the weak hadron current consists of the weak vector and weak
axial vector parts, \be
W^a_\mu(q)\,=\,J^a_\mu(q)\,+\,J^a_{\,5\mu}(q)\,,   \label{whc} \ee
and the momentum transfer $q=p_i-p_f=P_f-P_i$. In what follows, we
deal only with the weak axial hadron current $J^a_{\,5\mu}(q)$.

We define the potentials and quasipotentials in a similar way, as
it is done in Appendix A of Ref.\,\cite{ATA}.

We discuss next how the operator of a weak axial nuclear current
$j^a_{\,5\mu}$, used in conjunction with the equation describing
our NN system, can be related to the field--theoretical current
$J^a_{\,5\mu}(q)$. Let the time evolution of the NN system be
described by a Hamiltonian $H=T+V$, where $T$ is the kinetic
energy and $V$ is the nuclear potential, which we take as the sum
of the OBEPs, \be V\,=\,\sum_{B=\pi,\rho,\omega,a_1...}\,V_B\,.
\label{Vnuc} \ee If the nuclear physics calculations, based on the
current $j^a_{\,5\mu}$ and on the eigenfunctions of the
Hamiltonian $H$ should reflect the PCAC, then the current should
satisfy the continuity equation \be \vec q \cdot \vec
j^a_{\,5}(\vec q)\,=\,[\,H\,,\,\rho^a_{\,5}(\vec q)\,]\,+\,if_\pi
m^2_\pi \Delta^\pi_F(q^2)\,m^a(\vec q)\,,  \label{NCE} \ee where
$m^a$ is the associated pion absorption amplitude. Supposing that
the current consists of the one-- and two--nucleon components,
this equation splits into the following set of equations \bea \vec
q_i \cdot \vec j^a_{\,5}(1,\vec
q_i)\,&=&\,[\,T_i\,,\,\rho^a_{\,5}(1,\vec q_i)\,]\,+\,if_\pi
m^2_\pi
\Delta^\pi_F(q^2)\,m^a(1,\vec q_i)\,,\quad i=1,2\,,  \label{NCEoi}  \\
\vec q \cdot \vec j^a_{\,5}(2,\vec
q)\,&=&\,[\,T_1+T_2\,,\,\rho^a_{\,5}(2,\vec q)\,]\,+\,
([\,V\,,\,\rho^a_{\,5}(1,\vec q)\,]+\ot) \nonumber \\
&&+\,if_\pi m^2_\pi \Delta^\pi_F(q^2)\,m^a(2,\vec q)\,.
\label{NCEt} \eea In Eq.\,(\ref{NCEt}), we neglected
$\rho^a_{\,5}(2,\vec q)$ in the second commutator on the right
hand side. Taking into account that $q_0=q_{10}+q_{20}$, we find
that Eqs.\,(\ref{dWANEC}) and (\ref{NCEt}) are in the full
correspondence in the space of the nuclear states, that are
described by the eigenfunctions of the Hamiltonian $H$. So we can
consider the current $j^a_{\,5\mu,\,B}(2)$, defined in
Eq.\,(\ref{JnaB}), as the WANEC of the range B.

\section{The WANECs of the $\sigma$ meson range}
\label{happB}

Realistic OBEPs contain standardly the $\sigma$ meson exchange,
describing the attraction at medium distances. On the other hand,
our WANECs do not contain the component due to this exchange so
far. This is due to the fact that our Lagrangians reflect the
non--linear chiral symmetry and it is not clear how to include the
$\sigma$ meson into the scheme consistently. Here we construct the
axial exchange current of the $\sigma$ range starting from a
Lagrangian \be \Delta {\cal L}\,=\,g_\sigma {\bar
N}N\,\phi\,+\,i\frac{g_A}{f_\pi}g_\sigma\, {\bar N}\gamma_5(\vec
\tau \cdot \vec \pi)N\,\phi\,+\,ig_{\pi NN} {\bar N}\gamma_5(\vec
\tau \cdot \vec \pi)N\,+\,ig_A g_{\rho} {\bar
N}\gamma_\nu\gamma_5(\vec \tau \cdot {\vec a}_\nu )N\,, \label{DL}
\ee where besides the pseudoscalar $\pi NN$ coupling standardly
accepted $\sigma NN$ coupling is present. The relativistic
amplitudes derived from this Lagrangian by analogy with
Sect.\,\ref{CH1} for other exchanges of our model are the nucleon
Born amplitude $J^a_{\,5\mu,\,\sigma}$ and the only potential
contact term ${J}^a_{5\mu,\,c\,\sigma}(\pi)$. It can be verified
that these amplitudes satisfy the PCAC equation \be
q_\mu\left[\,{J}^a_{5\mu,\,\sigma}\,+\,{J}^a_{5\mu,\,c\,\sigma}\,\right]\,=\,
i f_\pi m^2_\pi\,\Delta^\pi_F(q^2)\,\left[\, M^a_\sigma\, +\,
M^a_{c\,\sigma}\,\right]\,.   \label{dJSEX} \ee Using the methods
developed in Sect.\,\ref{CH2}, we obtain the pair term of the
$\sigma$ range \bea {\vec j}^{\,\,a}_{5,\sigma}(pair)\,&=&\,i g_A
F_A \frac{g^2_\sigma}{(2M)^3}\, [\qb\times
(\Pa\,+\,\vq)\,+\,i\Pa\times(\sa\times\vq)]\,\ta\,
\Delta^\sigma_F(\qb^{\,\,2})\,+\,\ot\,,  \label{SCSPT}  \\
j^a_{50,\,\sigma}(pair) \,&=&\,g_A F_A \frac{g^2_\sigma}{(2M)^2}\,
(\sa\cdot(\Pa\,+\,\vq))\,\ta\,
\Delta^\sigma_F(\qb^{\,\,2})\,+\,\ot\,.  \label{TCSPT} \eea For
the transition $^{3}S_1-^{3}D_1\,\rightarrow\,^{1}S_0$, only the
second term at the right hand side of the space component of the
current ${\vec j}^{\,\,a}_{5,\sigma}(pair)$ contributes. However,
being proportional to $\vq$, it is of little importance at the
threshold. In  contrast to the current ${\vec A}_\pm (S)$
presented in Eq.\,(2.5a) of the Ref.\,\cite{TR}, our current
(\ref{SCSPT}) does not contain the spurious term  analogous to
$-{\vec \sigma}_1 {\vec k}^{\,\,2}/4$.

\end{document}